\documentclass[usenatbib]{mnras}
\usepackage{rotating}
\usepackage{colortbl}
\usepackage{color}
\usepackage[fleqn]{amsmath}
\usepackage{amssymb}
\usepackage{amsfonts}
\usepackage{amsmath}
\usepackage{verbatim}
\usepackage{scalefnt}
\usepackage{float}
\usepackage[percent]{overpic}

\def\lsim{\mathrel{\rlap{\lower 3pt \hbox{$\sim$}} \raise 2.0pt \hbox{$<$}}}
\def\gsim{\mathrel{\rlap{\lower 3pt \hbox{$\sim$}} \raise 2.0pt \hbox{$>$}}}

\newcommand{\comments}[1]{} 

\title[Shocks and angular momentum flips in mergers]{Shocks and angular momentum flips: a different path to feeding the nuclear regions of merging galaxies}

\author[P.~R. Capelo and M. Dotti]{Pedro R. Capelo$^{1}$\thanks{E-mail: pcapelo@physik.uzh.ch} and Massimo Dotti$^{2,3}$\\
$^1$Center for Theoretical Astrophysics and Cosmology, Institute for Computational Science, University of Zurich,\\
Winterthurerstrasse 190, CH-8057 Z$\ddot{u}$rich, Switzerland\\
$^2$Dipartimento di Fisica G. Occhialini, Universit$\grave{a}$ degli Studi di Milano Bicocca, Piazza della Scienza 3, I-20126 Milano, Italy\\
$^3$INFN, Sezione Milano--Bicocca, Piazza della Scienza 3, I-20126 Milano, Italy
}

\begin{document}

\maketitle

\begin{abstract}
We study the dynamics of galaxy mergers, with emphasis on the gas feeding of nuclear regions, using a suite of hydrodynamical simulations of galaxy encounters. The high spatial and temporal resolution of the simulations allows us to not only recover the standard picture of tidal-torque induced inflows, but also to detail another, important feeding path produced by ram pressure. The induced shocks effectively decouple the dynamics of the gas from that of the stars, greatly enhancing the loss of gas angular momentum and leading to increased central inflows. The ram-pressure shocks also cause, in many cases, the entire galactic gas disc of the smaller galaxy to abruptly change its direction of rotation, causing a complete ``flip'' and, several $10^8$~yr later, a subsequent ``counter-flip''. This phenomenon results in the existence of long-lived decoupled gas--stellar and stellar--stellar discs, which could hint at a new explanation for the origin of some of the observed kinematically decoupled cores/counter-rotating discs. Lastly, we speculate, in the case of non-coplanar mergers, on the possible existence of a new class of remnant systems similar to some of the observed X-shaped radio galaxies.\\
\end{abstract}



\section{Introduction}\label{angmomflips:sec:Introduction}

Gas inflows toward the central regions of disc galaxies, fuelling nuclear bursts of star formation and active galactic nuclei (AGN), have been ascribed to different physical mechanisms. Most of these require deviations from axial symmetry in the matter distribution (hence in the gravitational potential), in order to torque the interstellar medium (ISM), eroding the gas angular momentum budget. Well studied examples are galactic-scale stellar bars \citep[e.g.][]{Sand76, Rob79, Atha92}. As such large-scale structures can fail at fuelling the innermost regions of the host in presence of orbital resonances, additional non-axisymmetric features have been proposed, such as bar-driven nuclear spirals \citep{mac04a, mac04b, fanali15}, systems of nested bars \citep{shlo89, hel07} or, more generally, of nested non-axisymmetric structures \citep[e.g.][]{HQ2010}.

In such context, galaxy mergers occupy a unique niche, as in this case each galaxy torques the ISM of the companion from the largest to the smallest scales as the merger proceeds. Since the seminal works of, e.g., \citet{hernquist89}, \citet{Barnes91,BarnesHernquist96}, and \citet{MihosHernquist1996}, a coherent picture of the cause-effect chain connecting the large-scale merger to the gas inflow toward the two nuclei has been depicted: $(i)$ the tidal field of a galaxy promotes the formation of non-axisymmetric structures in the companion; $(ii)$ such internal features torque the gas, producing the gas inflow that fuels the host nuclei.

In this study, we analyse the results of the suite of high-resolution, $N$-body smoothed particle hydrodynamics (SPH) simulations of disc galaxy mergers presented in \citet{Capelo_et_al_2015}. The range of parameters studied was quite large, including several different values of initial mass ratios, orbital configurations, gas fractions, etc.

The exquisite mass and spatial resolution of the runs and the high time frequency of the simulation outputs allow us to recover the standard picture discussed above and, in addition, to study in great detail the non-negligible effect of the hydrodynamically driven shocks developing in the gas components of the two galaxies during their closest pericentric passages, as already discussed, for some configurations, in lower-resolution simulations by \citet{Barnes_2002}. The associated hydrodynamical torques result in a sharp and sudden variation of the gas angular momentum in the innermost regions of the two galaxies, both in {\it magnitude} and in {\it orientation} \citep[see also][]{Van_Wassenhove_et_al_2014,Capelo_et_al_2015}. The shocked material fuels the very central regions of the two galaxies, leading to the formation of dynamically decoupled structures between gas and stars, as well as (in some cases) between old and young stars. A fraction of the gas inflow has a sufficiently low angular momentum (with respect to the secondary nucleus in particular) to fuel a short burst of AGN activity and central star formation between pericentres, when the gas gets to the galactic nuclei. We anticipate, however, that the simulations analysed have a too coarse spatial resolution to confirm any gas inflow on (sub-)pc scales.

Explaining how gas feeds the nuclear regions of galaxies is not only crucial to the understanding of AGN activity and central star formation. Other, important physical phenomena can be better understood when one has a clearer picture of how the gas component behaves. For example, the above-mentioned formation of dynamically decoupled stellar structures hints at a new explanation for the origin of some of the kinematically decoupled components (or cores)/counter-rotating discs (hereafter, KDCs) observed in a variety of galaxies \citep[e.g.][]{Efstathiou_et_al_1982}, which is different from the usual explanation of retrograde merging orbits \citep[e.g.][]{Balcells_Quinn_1990}. Additionally, if one accepts the idea that the gas at $\sim$10--100-pc scales (which we do resolve) and (sub-)pc scales share the same direction of angular momentum and, moreover, assumes that the central BH's spin is influenced by the surrounding gas and determines the direction of (possible) radio jets, then some of the changes in the angular momentum in the non-coplanar mergers described in this paper could predict the existence of a new class of systems similar to some of the observed X-shaped radio galaxies \citep[XRGs;][]{Leahy_Parma_1992}.

The paper is organised as follows: in Section~\ref{angmomflips:sec:Numerical_Setup} we describe the main features of the simulation suite, in Section~\ref{angmomflips:sec:Results} we discuss the results in terms of the angular momentum evolution, whereas in Section~\ref{angmomflips:sec:Discussion_and_conclusions} we discuss our results, compare them with previous studies in the literature, and conclude.


\section{Numerical Setup}\label{angmomflips:sec:Numerical_Setup}

We simulate isolated mergers of disc galaxies, whose initial separation is equal to the sum of the two initial virial radii, set on initially parabolic orbits with a first pericentric distance equal to 20 per cent of the virial radius of the larger (hereafter: {\it primary}) galaxy.

Each galaxy is composed of a dark matter halo, described by a spherical Navarro--Frenk--White \citep{NFW1996} profile (with spin and concentration parameters 0.04 and 3, respectively) which exponentially decays beyond the virial radius \citep{Springel_White_1999}; a gaseous and stellar exponential disc with an isothermal sheet \citep{Spitzer_1942,Camm_1950}; a spherical \citet{Hernquist1990} stellar bulge; and a central supermassive black hole \citep[BH;][]{Bellovary10}. The mass fractions (with respect to the virial mass) for the galactic components are $4 \times 10^{-2}$ and $8 \times 10^{-3}$ for the disc and bulge, respectively.

The individual galactic angular momentum vector of each galaxy can initially be at an angle (0, $\pi$/4, or $\pi$ radians) with the global (orbital) angular momentum vector, making the mergers coplanar, prograde--prograde, prograde--retrograde, and retrograde--prograde; and non-coplanar. The smaller (hereafter: {\it secondary}) galaxy has a mass that is a fraction (1/10, 1/6, 1/4, 1/2, or 1) of the mass of the primary galaxy. The gas fraction in the disc can be 30 (standard) or 60 (high) per cent. The gravitational softening of the gas and stellar particles is 20 and 10~pc, respectively. The initial mass of the baryonic particles is $3.3 \times 10^3$~M$_{\odot}$ for stars and $4.6 \times 10^3$~M$_{\odot}$ for gas.

The primary galaxy in each merger has an initial virial mass of $2.21 \times 10^{11}$~M$_{\odot}$, stellar bulge mass of $1.77 \times 10^9$~M$_{\odot}$, and disc mass of $8.83 \times 10^9$~M$_{\odot}$. The initial disc scale radius of the merging galaxies varies from 1.13~kpc for the primary galaxy to 0.53~kpc for the secondary galaxy in the 1:10 merger (the initial disc scale height and bulge scale radius are 10 and 20 per cent, respectively, of the disc scale radius).

A total of 14 merger simulations were performed, using the $N$-body SPH code {\scshape gasoline} \citep{wadsley04} -- an extension of the $N$-body tree code {\scshape pkdgrav} \citep{stadel01} -- which includes line cooling for atomic hydrogen and helium, and metals \citep{Shen_Wadsley_Stinson_2010}, prescriptions for star formation (in which stars colder than 6000~K and denser than 100 a.m.u.~cm$^{-3}$ can stochastically form with a star formation efficiency of 0.015), supernova feedback, and stellar winds \citep{Stinson2006}, and a recipe for BH accretion and feedback \citep{Bellovary10}.

In addition to the simulations discussed in \citet{Capelo_et_al_2015} -- see their Tables~1 and 2 for a summary of the parameters described here -- we included a new simulation, where we have re-run one of the 1:2 coplanar, prograde--prograde mergers with the mass of BHs increased by a factor of 2.5 (changing the BH mass fraction from $1.6 \times 10^{-5}$ to $4 \times 10^{-5}$) with respect to the standard value from the BH-bulge mass relation \citep{MarconiHunt2003}, since it was recently found that such relation could evolve with redshift \citep{Merloni2010}. We note, however, that the specific implementation of BH physics (accreting versus non-accreting BHs; level of BH feedback efficiency; initial BH mass) has a negligible effect on the phenomena described in this paper, as it will be shown in Section~\ref{angmomflips:sec:The_other_mergers}.

All the above properties and sub-grid models were inherited from \citet{Capelo_et_al_2015}. We stress, however, that, for the specific purposes of this study, the implementation of star formation (and related feedback) is mostly useful only for the study of stellar--stellar decoupled discs (see Sections~\ref{angmomflips:sec:1to4coppropro} and \ref{angmomflips:sec:The_other_mergers}). However, the really essential aspect of this numerical investigation, in order to determine the fate of the gas, is the relatively high resolution: for example, the primary galaxy's disc scale radius is $\sim$60 times the gravitational softening of the gas particles.


\section{Results}\label{angmomflips:sec:Results}

\begin{figure}
\centering
\vspace{2.5pt}
\includegraphics[trim = 11mm 14mm 0mm 7mm, clip, width=1.20\columnwidth,angle=0]{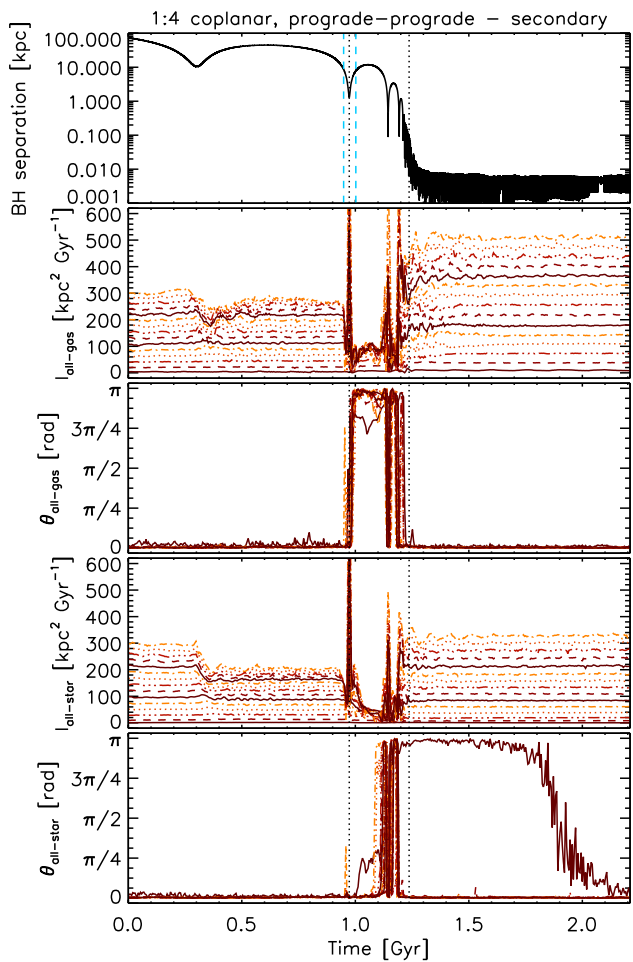}
\vspace{-5pt}
\caption[]{Temporal evolution of the specific angular momentum for the secondary galaxy in the 1:4 coplanar, prograde--prograde merger. In all panels, the vertical, dotted, black lines show the separation between the stochastic, the merger, and the remnant stage. The vertical, dashed, cyan lines in the first panel refer to the first and last row of Fig.~\ref{angmomflips:fig:density_snapshots}. First panel: separation between the two BHs. Second panel: magnitude of the gas specific angular momentum $l$ in concentric shells of 100-pc thickness around the local centre of mass near the secondary BH, in the inner 3~kpc (we only show half of the shells for clarity); $l$ grows monotonically with radius at the beginning and at the end of the run. Third panel: same as the second panel, but for the polar angle of the specific angular momentum vector. Fourth panel: same as the second panel, but for the stars. Fifth panel: same as the fourth panel, but for the polar angle.}
\label{angmomflips:fig:m4_hr_gf0_3_BHeff0_001_phi000000_angular_momentum_secondary_allgas_allstars_3kpc}
\end{figure}

In this section, we first present the results of one of the mergers in the suite (the 1:4 coplanar, prograde--prograde merger), to describe in detail the occurrence of shock-induced gas inflows and angular momentum flips. We then consider all the mergers as a whole, to discuss trends across the entire suite. We additionally describe separately two mergers, the 1:1 merger and the 1:2 inclined-primary merger, which present peculiar behaviours that could explain the existence of systems such as KDCs and XRGs.

Since angular momentum is a vectorial quantity which depends on the chosen coordinates system, we here define the centre of each galaxy as the local centre of mass around the central BH of the galaxy, calculated iteratively within a 100-pc sphere until convergence is achieved (i.e. when the fractional difference between the `new' and `old' local centre of mass is less than $10^{-4}$; \citealt{Capelo_et_al_2015}).


\subsection{The 1:4 coplanar, prograde--prograde merger}\label{angmomflips:sec:1to4coppropro}

We consider the 1:4 coplanar, prograde--prograde merger (also referred to as the {\it default merger}) as the representative example of all systems where shock-induced gas inflows and angular momentum flips occur, both because its mass ratio is at the boundary between major and minor mergers \citep{Mayer2013}, and because its other properties (BH growth, AGN activity, star formation, etc.) have already been thoroughly described in recent work \citep{Capelo_et_al_2015, Volonteri_et_al_2015a, Volonteri_et_al_2015b, Gabor_et_al_2016}, allowing for an easier comparison.

Fig.~\ref{angmomflips:fig:m4_hr_gf0_3_BHeff0_001_phi000000_angular_momentum_secondary_allgas_allstars_3kpc} shows the evolution of the secondary galaxy of the default merger\footnote{As the primary galaxy is significantly less perturbed by the interaction, we present the evolution of its angular momentum in Appendix~\ref{angmomflips:sec:Additional_figures} (see Fig.~\ref{angmomflips:fig:m4_hr_gf0_3_BHeff0_001_phi000000_angular_momentum_primary_allgas_allstars_3kpc}).}. The top panel shows the BH separation as a function of time. As in \citeauthor{Capelo_et_al_2015} (\citeyear{Capelo_et_al_2015}, \citeyear{Capelo_et_al_2016}), we divide the history of the encounter in three main stages, highlighted by the two dotted, vertical lines: the {\it stochastic} (or early) stage, from the beginning of the simulation to the second pericentric passage; the proper {\it merger stage}; and the {\it remnant} (or late) stage, when a merger remnant has formed. The second panel shows the magnitude ($l$) of the specific angular momentum of all the gas in spherical shells of thickness 100~pc, centred on the secondary galaxy's centre, within the inner 3~kpc. The choice for the definition of stages is now clear: the $l$-curves are more or less ``flat'' during the stochastic and remnant stages, with the angular momentum of the outer regions ($r \gtrsim 1$~kpc) only being perturbed by the tidal field of the companion during the first pericentric passage, whereas the $l$-curves undergo dramatic oscillations during the merger stage. The third panel shows the polar angle ($\theta$) of the angular momentum of all the gas in the same shells shown in the second panel. Once again, the three stages are clearly distinguishable: during the stochastic and remnant stages, the polar angle is $\sim$0, whereas during most of the merger stage the polar angle is $\sim$$\pi$~radians. The rotation of the gas around the secondary's centre abruptly changes sign (``flips'') at a time close to the second pericentric passage, changing the merger from prograde--prograde to prograde--retrograde, and then, after $\sim$200~Myr, changes sign again (``counter-flips''), towards the end of the merger stage, when the merger remnant is being formed. Note that the flip is not caused by the fact that the two gas discs overlap and we are therefore including counter-rotating primary gas particles in the calculation of the angular momentum: the same result applies when we consider only the gas particles that were part of the secondary galaxy at the beginning of the simulation\footnote{On the other hand, the sharp peaks in the angular momentum magnitude and the associated fast fluctuations in $\theta$ observed at the second, third and fourth pericentric passages are simply due to the presence of primary-gas particles within $\lsim 3$~kpc from the secondary galaxy's centre.}.

The physical explanation for the flip is highlighted in Fig.~\ref{angmomflips:fig:density_snapshots}, where we show the gas and stellar density of the default merger at around the time of the second pericentric passage (rows~1--8: times 0.948--1.002~Gyr). At this time, the separation between the two galactic centres is as low as $\sim$1~kpc (row~4 of Fig.~\ref{angmomflips:fig:density_snapshots}), so that the two galaxies actually cross each other. While the stellar and dark matter components are subject only to the gravitational interaction, the gas components experience a huge hydrodynamical torque as soon as the two galaxies impact. The ram pressure exerted by the primary gas onto the secondary gas is clearly observable before the second pericentre (as a straight discontinuity in the gas distribution\footnote{Note that such structure cannot be ascribed to the tidal field of the primary galaxy, as such field tends to elongate the galaxy in the perpendicular direction, as observable in the stellar density field (see the right column of Fig.~\ref{angmomflips:fig:density_snapshots}).}; see the left and central columns of row~2 of Fig.~\ref{angmomflips:fig:density_snapshots}) and dramatically alters the motion of the gas during the pericentric passage, clearly decoupling the dynamics of the gas from that of the stars. Such decoupling results in the formation of a bridge between the two galaxies, already observed by \citet{Barnes_2002} and reminiscent of the structures observed in deeply interpenetrating mergers \citep{Condon_et_al_1993,Gao_et_al_2003}.

\begin{figure*}
\centering
\vspace{0pt}
\begin{overpic}[width=0.65\columnwidth,angle=0]{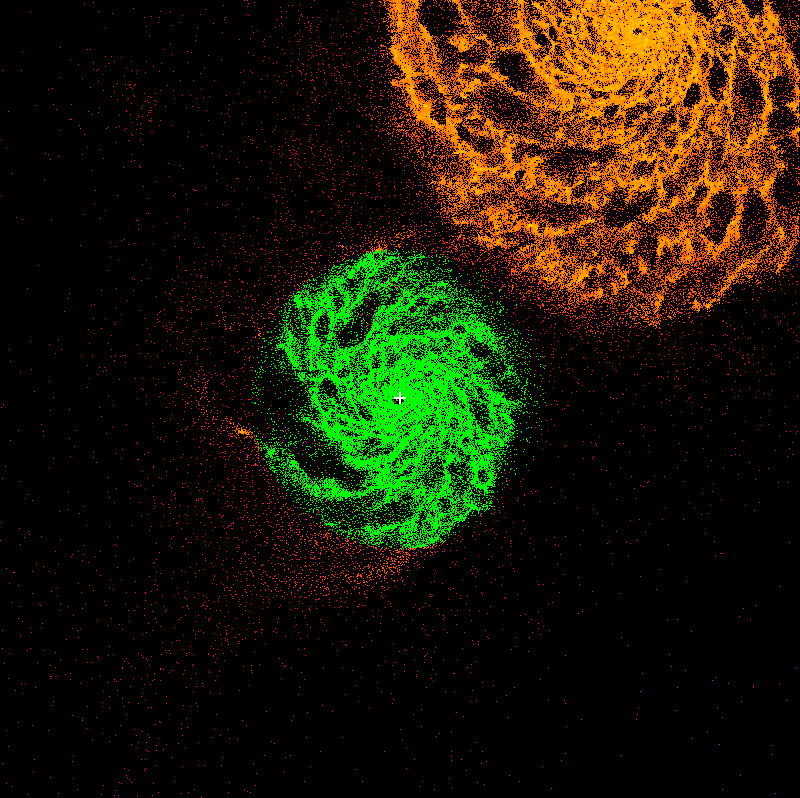}
\put (2,93) {\textcolor{white}{1}}
\end{overpic}
\begin{overpic}[width=0.65\columnwidth,angle=0]{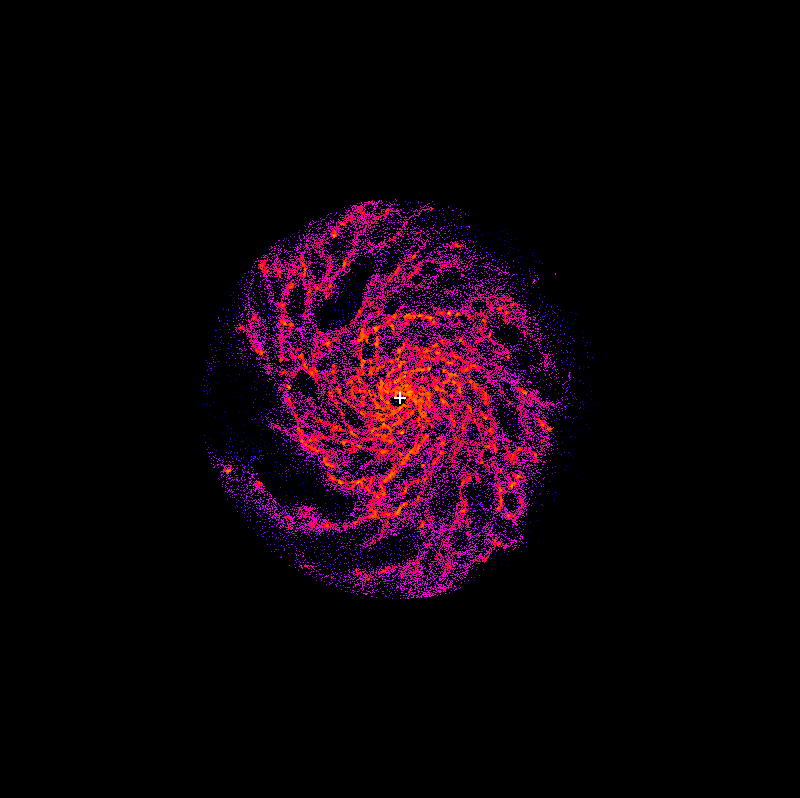}
\end{overpic}
\begin{overpic}[width=0.65\columnwidth,angle=0]{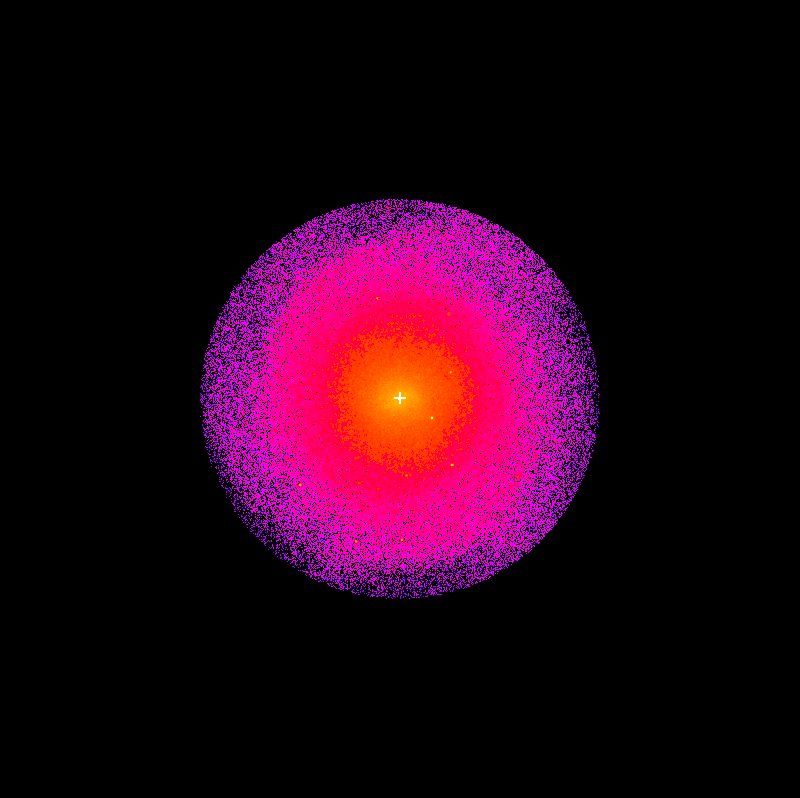}
\put (73,93) {\textcolor{white}{0.948~Gyr}}
\end{overpic}
\vskip 0.8mm
\begin{overpic}[width=0.65\columnwidth,angle=0]{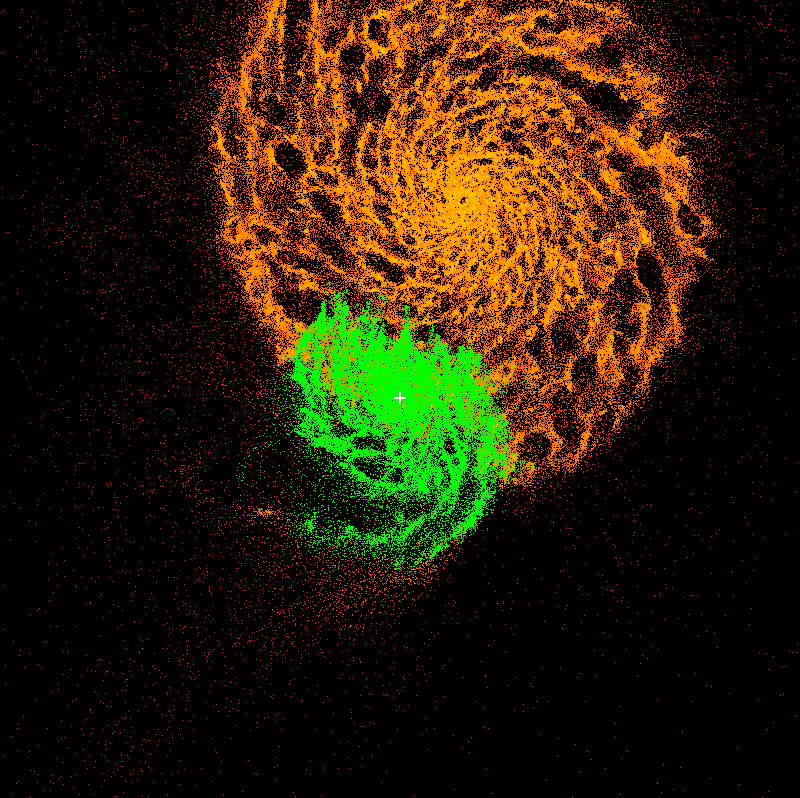}
\put (2,93) {\textcolor{white}{2}}
\end{overpic}
\begin{overpic}[width=0.65\columnwidth,angle=0]{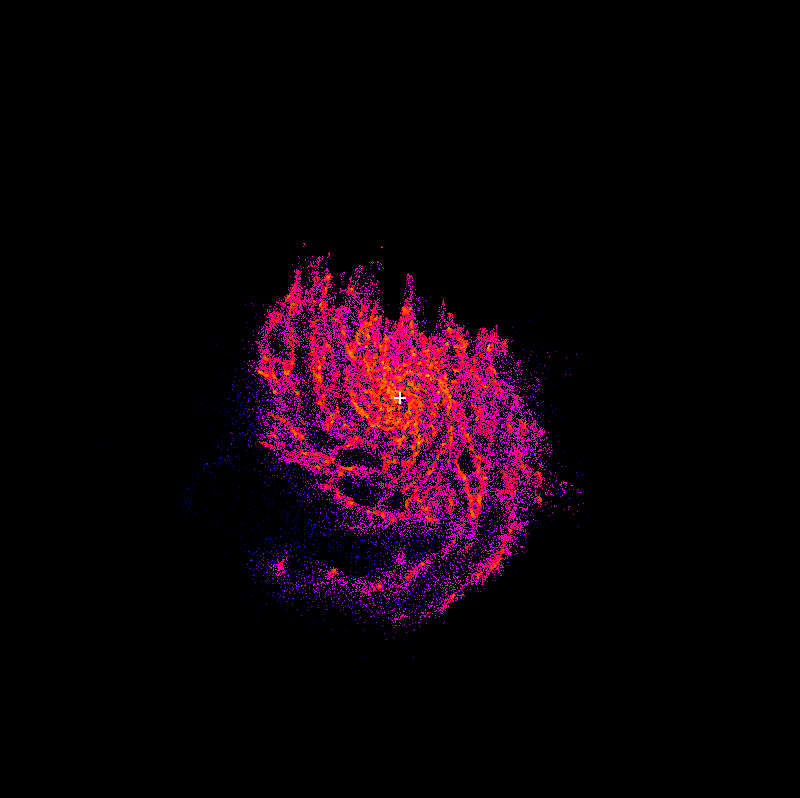}
\end{overpic}
\begin{overpic}[width=0.65\columnwidth,angle=0]{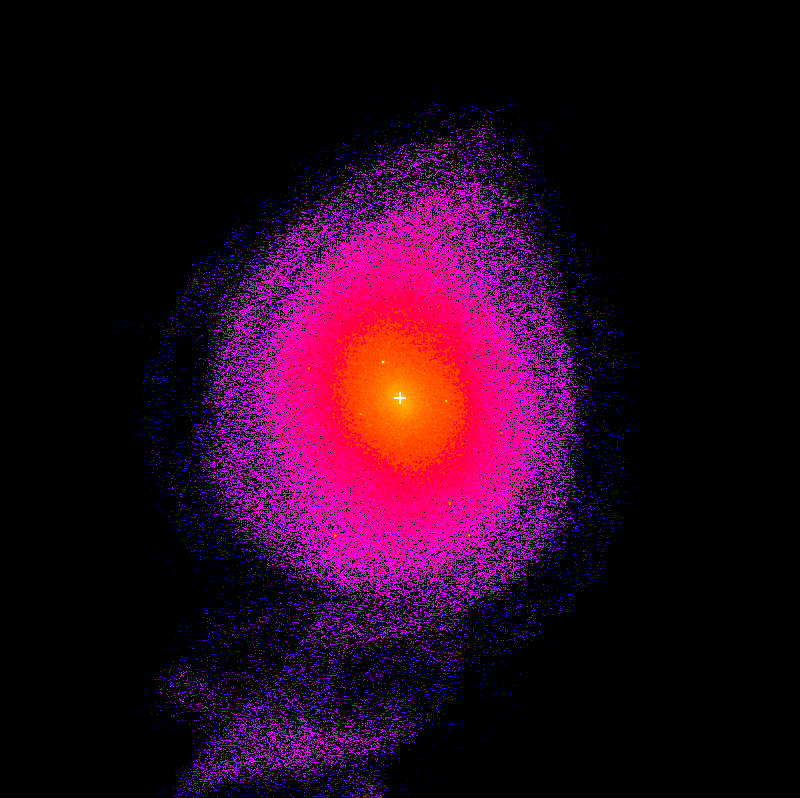}
\put (73,93) {\textcolor{white}{0.963~Gyr}}
\end{overpic}
\vskip 0.8mm
\begin{overpic}[width=0.65\columnwidth,angle=0]{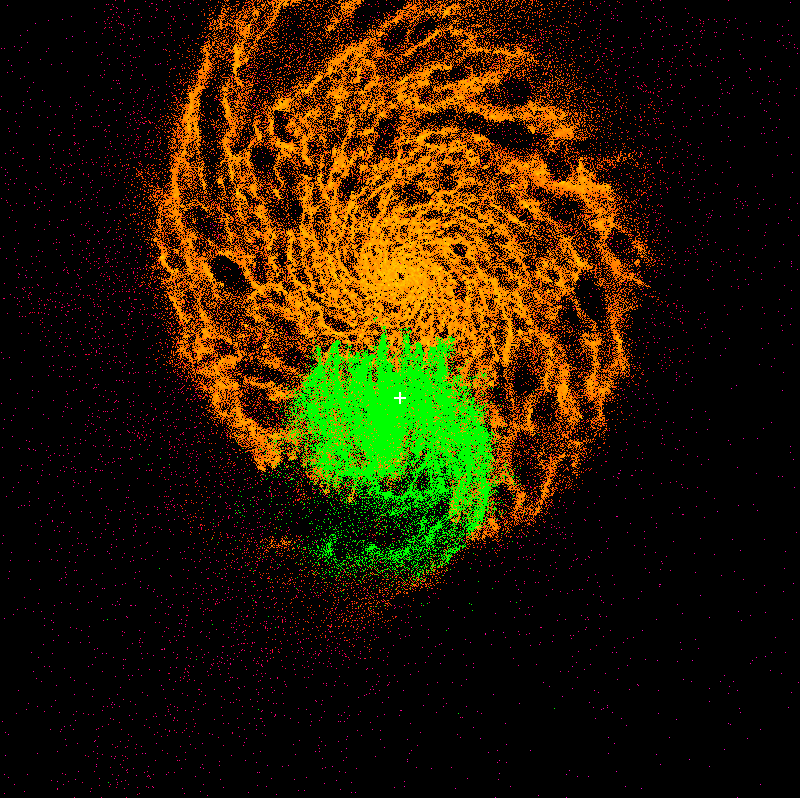}
\put (2,93) {\textcolor{white}{3}}
\end{overpic}
\begin{overpic}[width=0.65\columnwidth,angle=0]{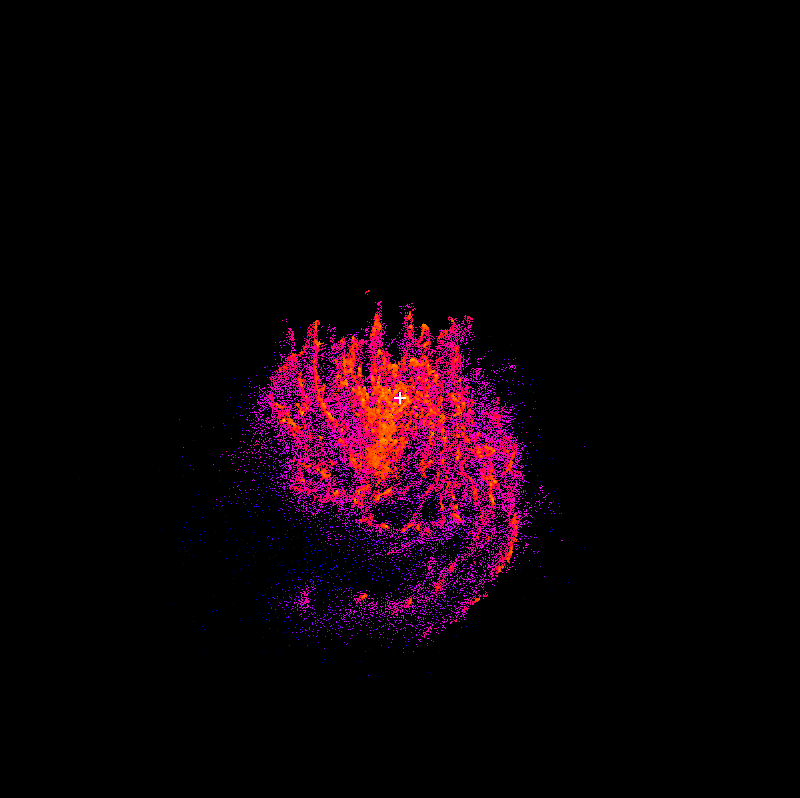}
\end{overpic}
\begin{overpic}[width=0.65\columnwidth,angle=0]{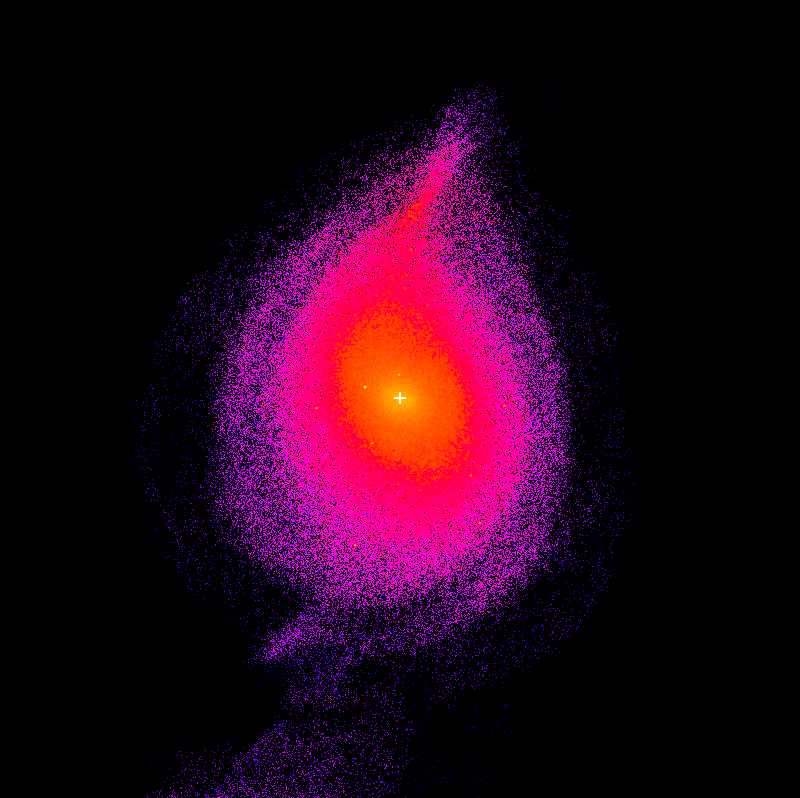}
\put (73,93) {\textcolor{white}{0.968~Gyr}}
\end{overpic}
\vskip 0.8mm
\hspace{0.0pt}\begin{overpic}[width=0.65\columnwidth,angle=0]{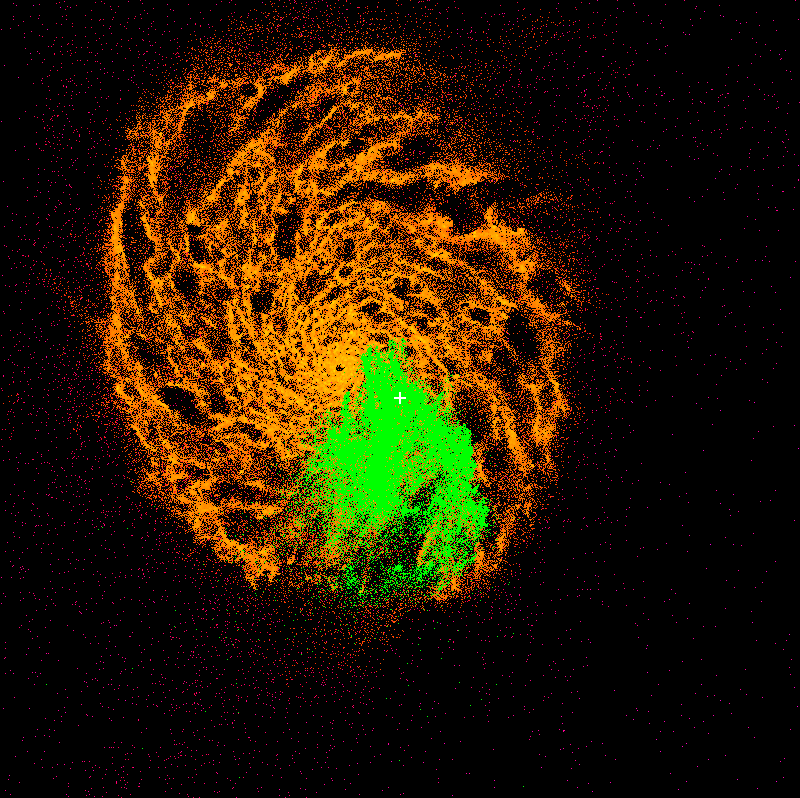}
\put (2,93) {\textcolor{white}{4}}
\end{overpic}
\begin{overpic}[width=0.65\columnwidth,angle=0]{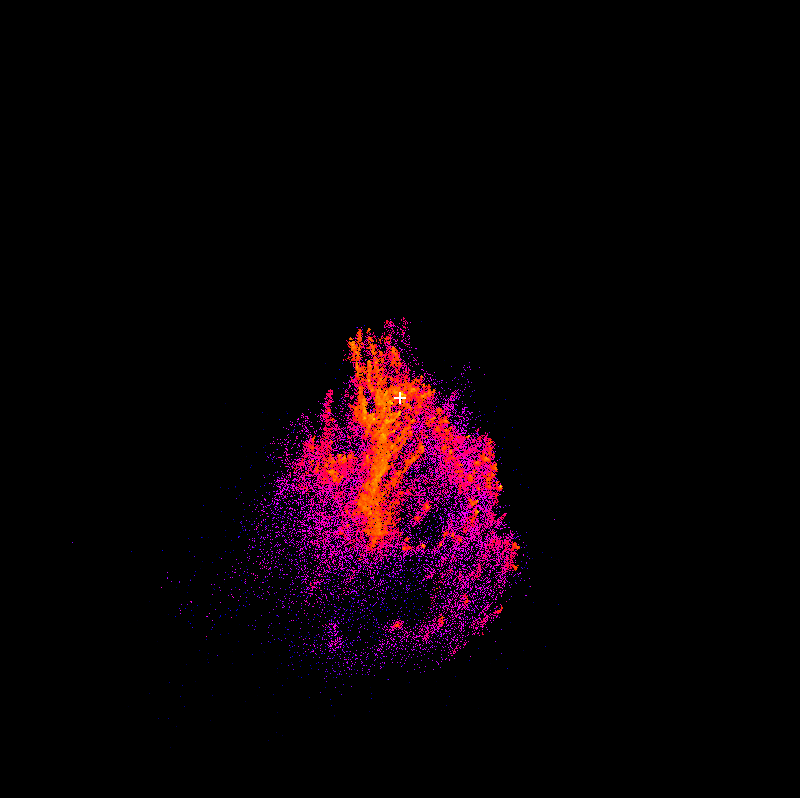}
\put (29,93) {\textcolor{white}{Second pericentre}}
\end{overpic}
\begin{overpic}[width=0.65\columnwidth,angle=0]{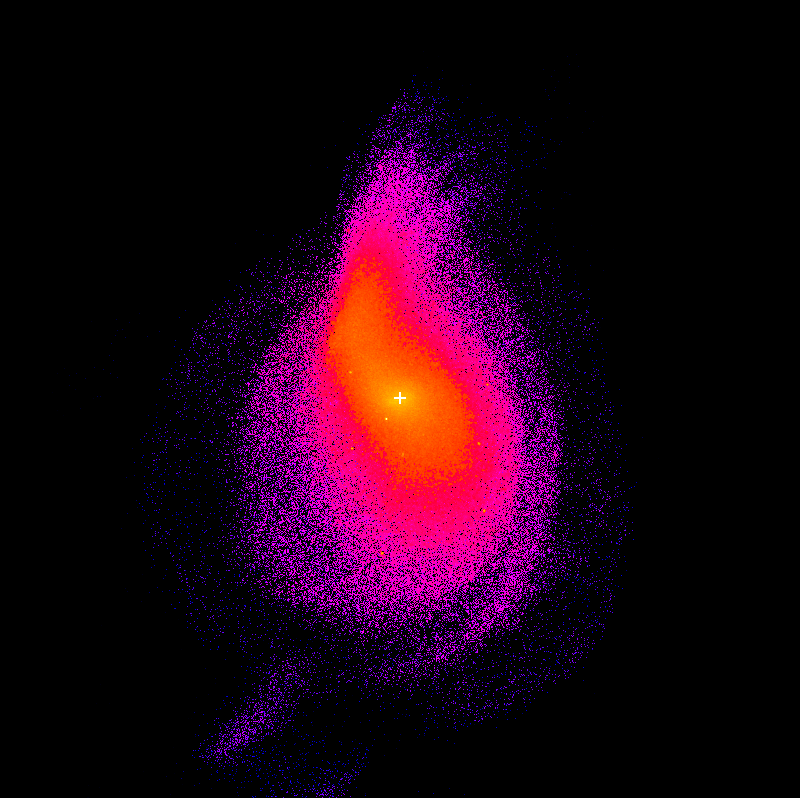}
\put (73,93) {\textcolor{white}{0.973~Gyr}}
\end{overpic}
\vskip 0.8mm
\hspace{161.0pt}\begin{overpic}[height=0.6cm,width=1.31\columnwidth,angle=0]{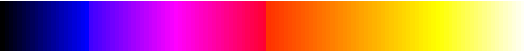}
\put (1,2) {\textcolor{white}{$10^1$}}
\put (23,2) {$10^3$}
\put (48,2) {$10^5$}
\put (72,2) {$10^7$}
\put (95,2) {$10^9$}
\end{overpic}
\vspace{0pt}
\caption[Density snapshots]{Detailed evolution of the 1:4 coplanar, prograde--prograde merger close to the second pericentre. Left column: positions of the gas particles of both galaxies, with the gas originally within 3~kpc from the secondary galaxy's centre marked in green. Central column: density map of the gas originally within 3~kpc from the centre of the secondary. Right column: same as the central column, but for the stars. The colour bar shows the (logarithmic) density scale in units of $2.2 \times 10^5$~M$_{\odot}$~kpc$^{-3}$. {\it -- continues on the next page --}}
\label{angmomflips:fig:density_snapshots}
\end{figure*}
 
\begin{figure*}
\centering
\vspace{0pt}
\begin{overpic}[width=0.65\columnwidth,angle=0]{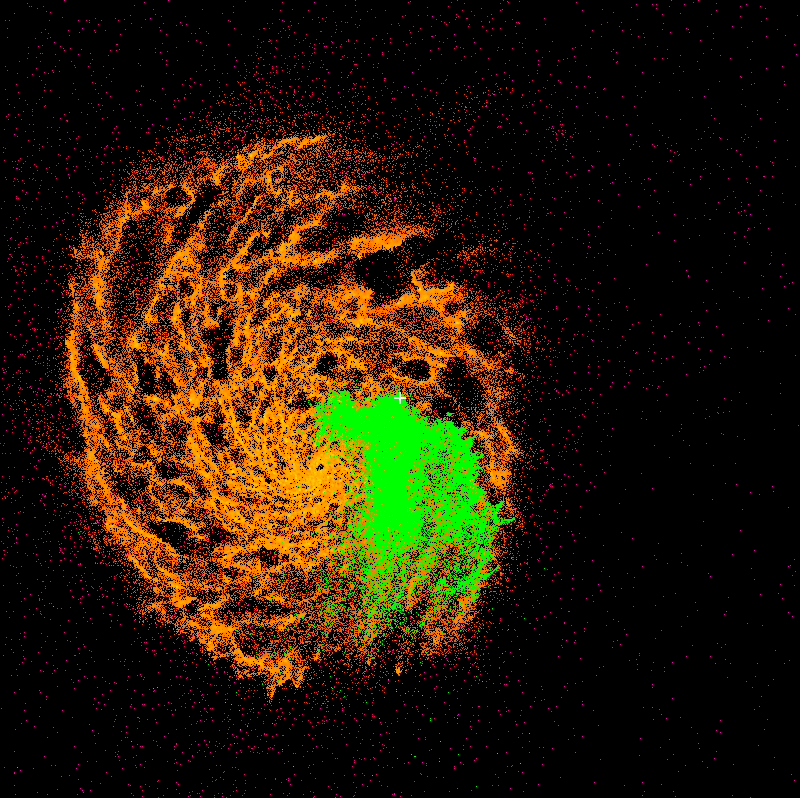}
\put (2,93) {\textcolor{white}{5}}
\end{overpic}
\begin{overpic}[width=0.65\columnwidth,angle=0]{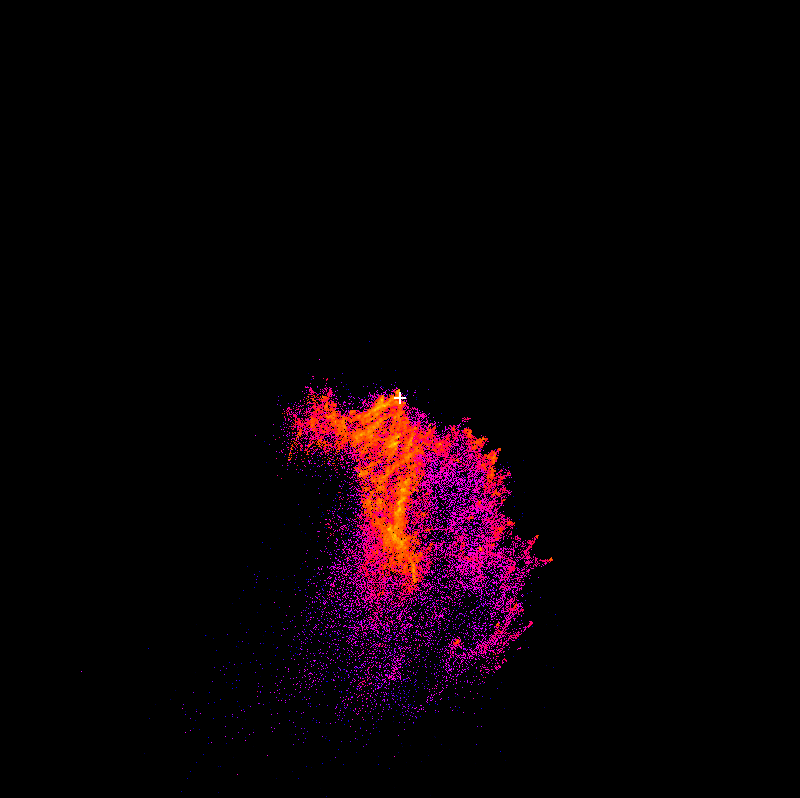}
\end{overpic}
\begin{overpic}[width=0.65\columnwidth,angle=0]{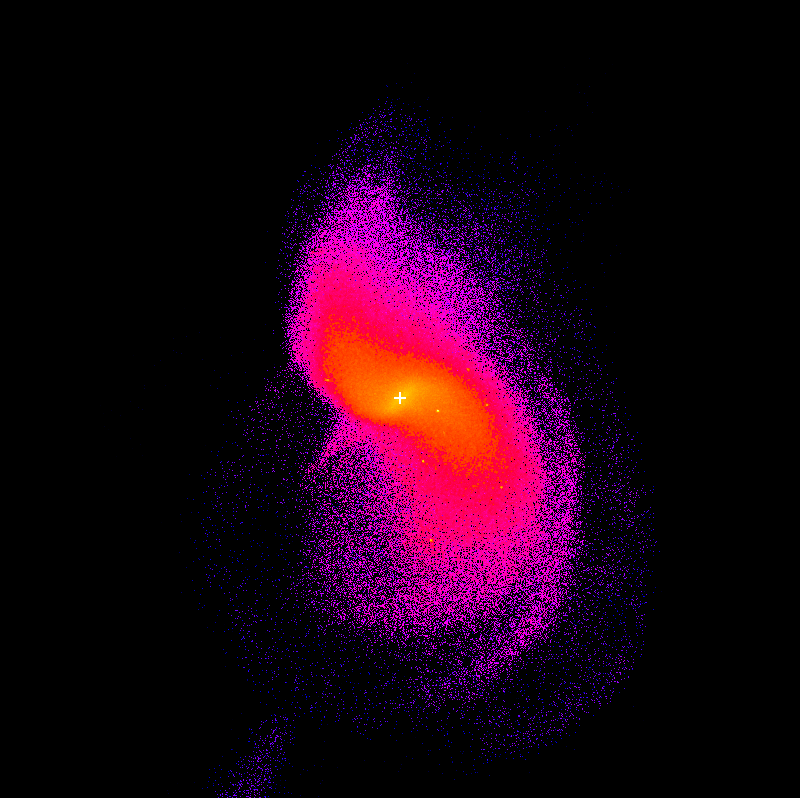}
\put (73,93) {\textcolor{white}{0.978~Gyr}}
\end{overpic}
\vskip 0.8mm
\begin{overpic}[width=0.65\columnwidth,angle=0]{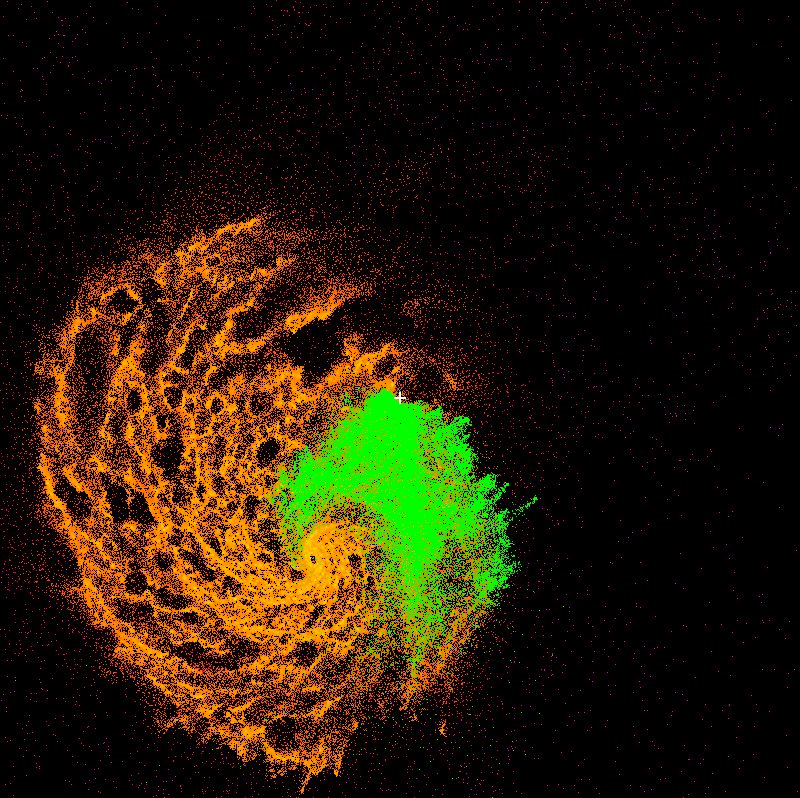}
\put (2,93) {\textcolor{white}{6}}
\end{overpic}
\begin{overpic}[width=0.65\columnwidth,angle=0]{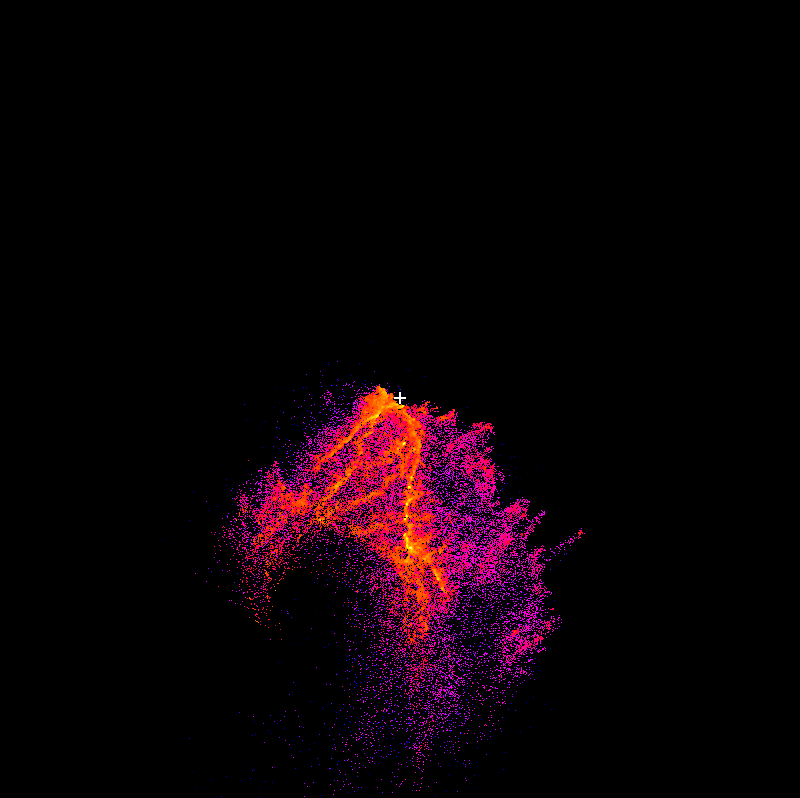}
\end{overpic}
\begin{overpic}[width=0.65\columnwidth,angle=0]{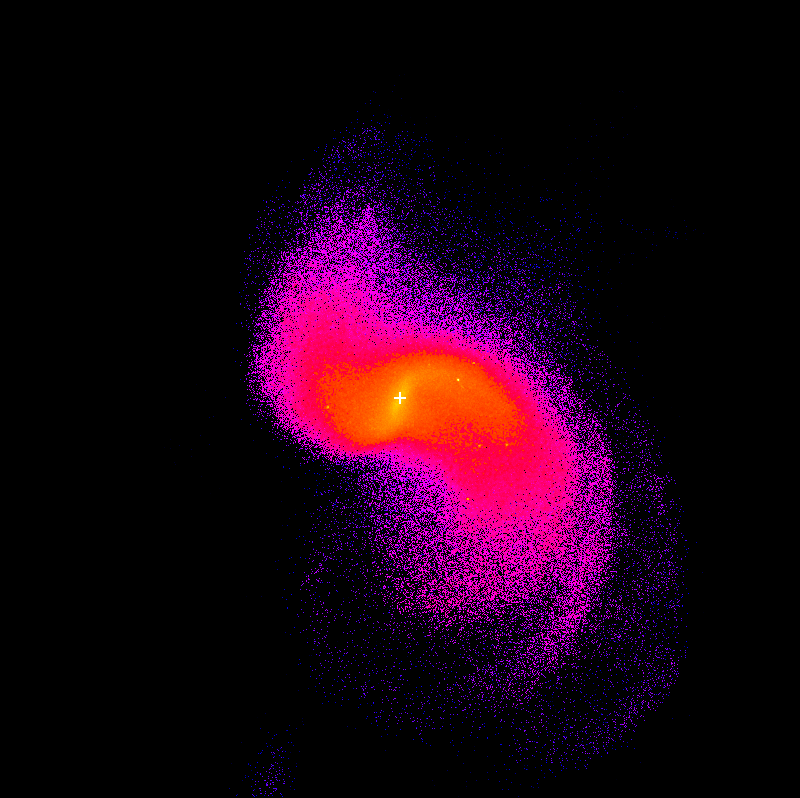}
\put (73,93) {\textcolor{white}{0.983~Gyr}}
\end{overpic}
\vskip 0.8mm
\begin{overpic}[width=0.65\columnwidth,angle=0]{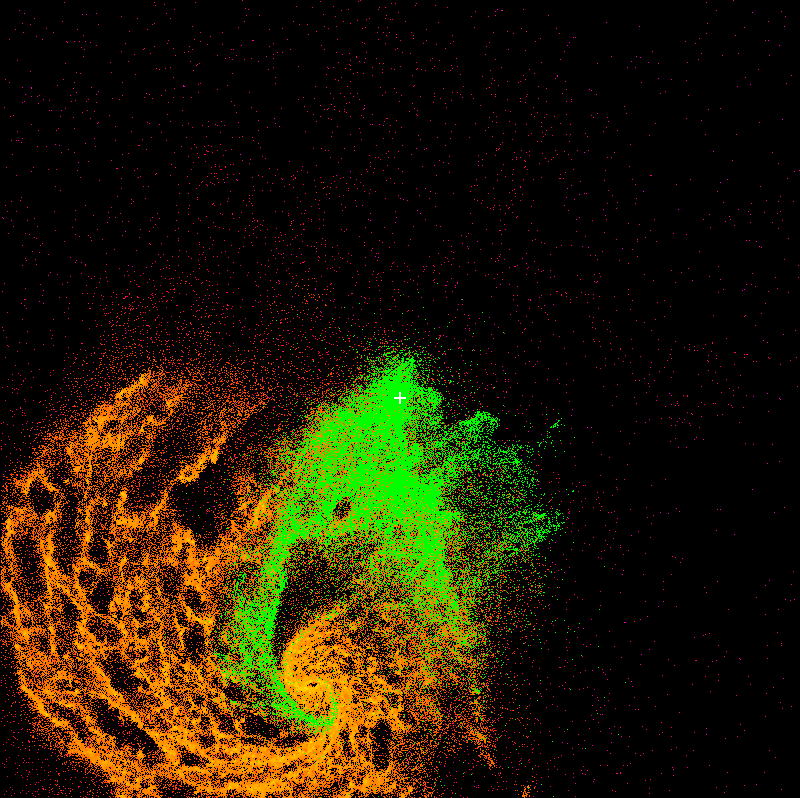}
\put (2,93) {\textcolor{white}{7}}
\end{overpic}
\begin{overpic}[width=0.65\columnwidth,angle=0]{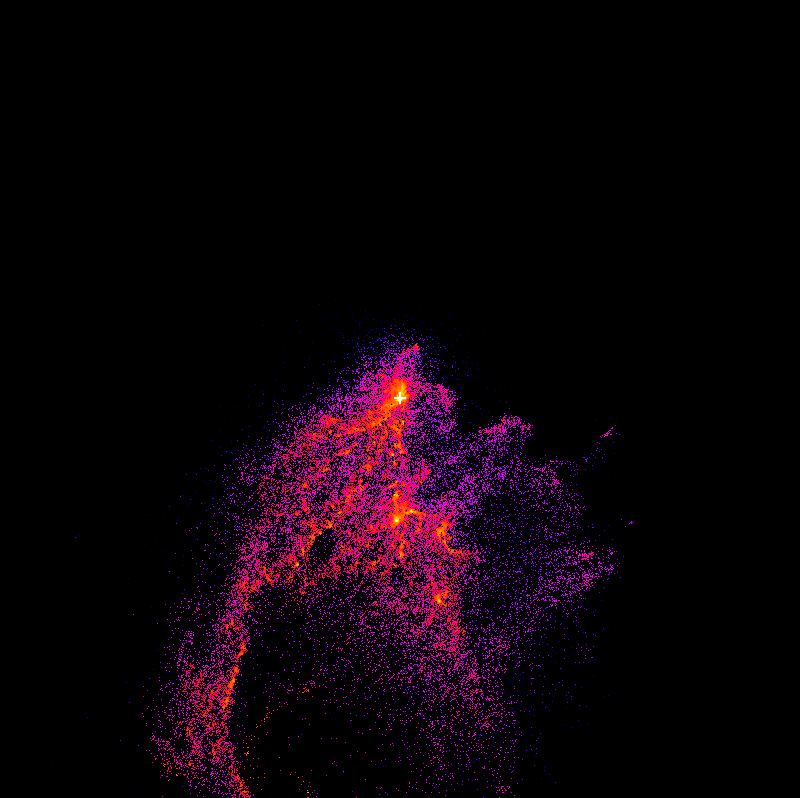}
\end{overpic}
\begin{overpic}[width=0.65\columnwidth,angle=0]{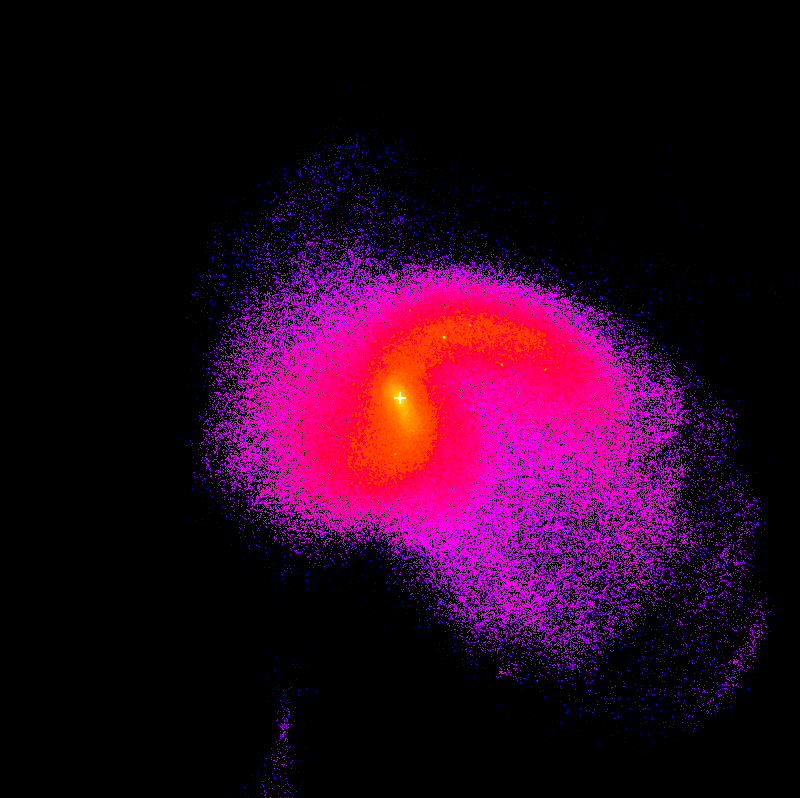}
\put (73,93) {\textcolor{white}{0.992~Gyr}}
\end{overpic}
\vskip 0.8mm
\hspace{0.0pt}\begin{overpic}[width=0.65\columnwidth,angle=0]{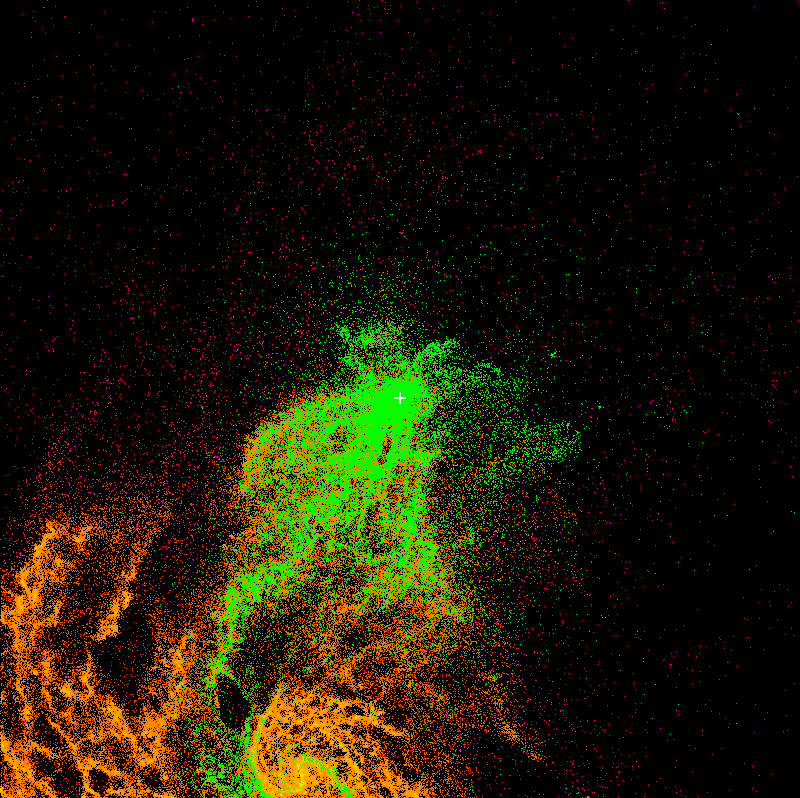}
\put (2,93) {\textcolor{white}{8}}
\end{overpic}
\begin{overpic}[width=0.65\columnwidth,angle=0]{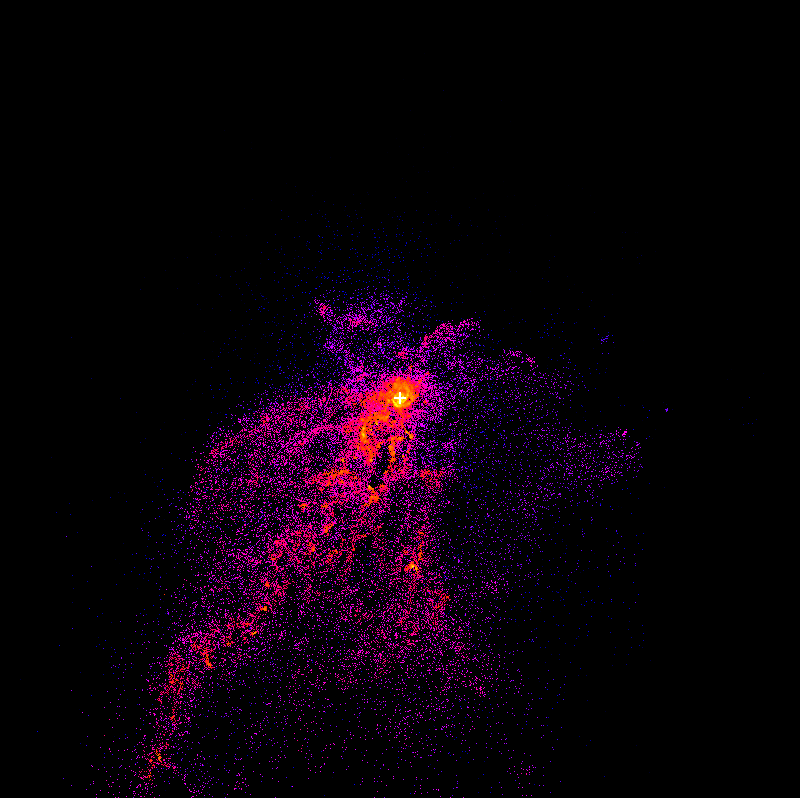}
\end{overpic}
\begin{overpic}[width=0.65\columnwidth,angle=0]{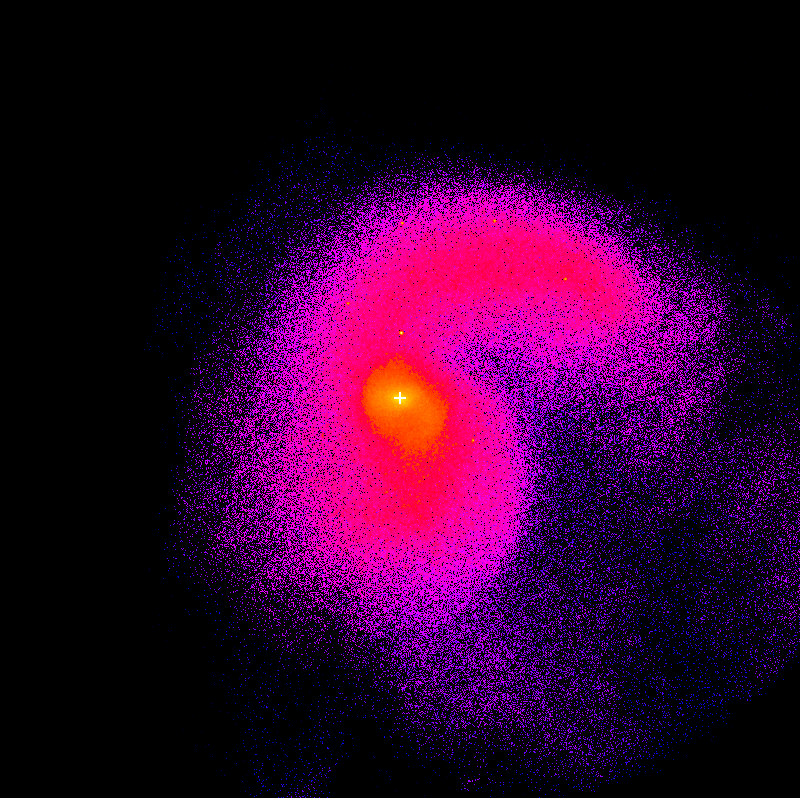}
\put (73,93) {\textcolor{white}{1.002~Gyr}}
\end{overpic}
\vskip 0.8mm
\hspace{174.0pt}\begin{overpic}[height=0.6cm,width=1.31\columnwidth,angle=0]{inpaper/density_maps/colourbar.png}
\put (1,2) {\textcolor{white}{$10^1$}}
\put (23,2) {$10^3$}
\put (48,2) {$10^5$}
\put (72,2) {$10^7$}
\put (95,2) {$10^9$}
\end{overpic}
\vspace{0pt}
\contcaption{{\it from the previous page --} The secondary galaxy's centre is marked with a white cross. The size of the left panels is 16$\times$16~kpc, whereas the central and right panels are 12$\times$12~kpc in size, to better detail the gaseous and stellar structures within the secondary. The times of the first and last row are marked by the vertical, dashed, cyan lines in Fig.~\ref{angmomflips:fig:m4_hr_gf0_3_BHeff0_001_phi000000_angular_momentum_secondary_allgas_allstars_3kpc}.}
\end{figure*}

After the second pericentre (rows 5--8 of Fig.~\ref{angmomflips:fig:density_snapshots}), the gas dynamics is once again dominated by gravity. The gas that suffered the strongest hydrodynamical deceleration (the first impacting onto the primary, see again row~2 of Fig.~\ref{angmomflips:fig:density_snapshots}) un-binds from its host, and accretes onto the primary onto a spiral wake clearly visible in rows 6--8 of Fig.~\ref{angmomflips:fig:density_snapshots}. Most of the gas, however, remains bound to the secondary. Because of the same hydrodynamical deceleration, the gas on the right side of the secondary does not manage to overcome the galaxy centre on the right, as required by the original galaxy rotation, but passes below the secondary centre (see especially row 5 of Fig.~\ref{angmomflips:fig:density_snapshots}) and starts rotating clockwise\footnote{The gas on the left side, on the other hand, creates a dense column of gas, clearly visible in rows 4--6 of Fig.~\ref{angmomflips:fig:density_snapshots}, which does not move much with respect to the secondary's centre and therefore does not contribute significantly to the angular momentum's budget.}.

The counter-flip, observed at the end of the merger stage, has a similar explanation. As the separation between the two galaxies drops below 1~kpc, the almost unperturbed gaseous disc of the primary and the secondary counter-rotating one shock one onto the other, and the secondary gas component is dragged by the more massive primary component and forced to co-rotate with it. The counter-flip does not occur at the third pericentric passage, as one could expect, because the secondary disc, during the merger stage, has become more compact (see row~8 of Fig.~\ref{angmomflips:fig:density_snapshots}) and therefore more resilient against ram pressure from the primary. The moment at which the counter-flip occurs is likely the result of a combination of how compact the secondary disc has become and how close the two galactic centres are during late pericentric passages (see also Section~\ref{angmomflips:sec:The_other_mergers}).

The evolution of the magnitude and polar angle of the stellar specific angular momentum (fourth and fifth panels in Fig.~\ref{angmomflips:fig:m4_hr_gf0_3_BHeff0_001_phi000000_angular_momentum_secondary_allgas_allstars_3kpc}) reinforces the picture described above. Stars start with the same specific angular momentum of the gas (by construction), except in the inner kpc where, due to the presence of a bulge, the stellar angular momentum is slightly lower than that of the gas. The first drop -- at the first pericentric passage -- in the angular momentum magnitude of the stars in the outermost shells is very similar to that of the gas, as the dynamics of both components is perturbed by the tidal perturbation of the primary galaxy\footnote{After the first pericentric passage, however, the (dissipative) gas component regains almost all its specific angular momentum, whereas the (dissipationless) stars do not.}. The drop of stellar specific angular momentum at the second pericentric passage happens on a longer timescale with respect to its gaseous counterpart\footnote{The narrow peaks in $l$ and $\theta$ have the same explanation of those observed in the gas component; see Footnote~2.}. This is due to the fact that the stellar component is not affected by the sudden braking force exerted onto the gas. However, new stars formed from the gas that has flipped during the merger stage retain $\theta \sim \pi$~radians. Since this recent stellar population remains sub-dominant almost everywhere in the galaxy, the stellar polar angle of the outer shells remains almost unperturbed. The exception is the central 100-pc region of the secondary galaxy, where the low angular momentum gas is focussed, and where the burst of star formation is extremely significant \citep{Capelo_et_al_2015,Volonteri_et_al_2015a,Volonteri_et_al_2015b}. Within these central $\sim$100~pc, the stellar distribution starts from $\theta \sim 0$ and gradually grows to $\sim$$\pi$~radians as the angular momentum of the newly formed stars becomes more and more important with respect to that of the pre-second-pericentre population.

The situation does not change for another $\sim$0.6~Gyr, until the time when the very recent stars, formed from the gas that has already counter-flipped, start becoming predominant and bring the polar angle back to $\sim$0. The reason why the decrease starts with so much delay after the gas has counter-flipped (compared to the almost immediate beginning of the increase after the gas has flipped) is because star formation is much higher during the merger stage than during the remnant stage (see \citealt{Capelo_et_al_2015}). The net result is the long-lived existence of a KDC within the merger remnant, and we will discuss the implications of this occurrence in Section~\ref{angmomflips:sec:Discussion_and_conclusions}.

At the end of the simulation, the angular momentum of the merger remnant is quite similar to that of the (original) primary galaxy (see Fig.~\ref{angmomflips:fig:m4_hr_gf0_3_BHeff0_001_phi000000_angular_momentum_primary_allgas_allstars_3kpc}), with the gas having almost the same specific angular momentum it had in the larger galaxy, and the stars having lost instead a significant part (this is due to the partial destruction of the two merging stellar discs and the build-up of a larger remnant bulge).


\subsection{General trends}\label{angmomflips:sec:The_other_mergers}

The phenomenon described in the previous section, namely the shock-induced gas inflows and angular momentum flips at the second pericentric passage, is not unique to the default merger. In this section, we identify the general trends found across the entire set of encounters. In the online-only material we show the history of the gaseous and stellar angular momenta of all the mergers not shown in the paper.

\subsubsection{General behaviour}\label{angmomflips:sec:General_behaviour}

{\it Early stages:} in {\it all the systems} in the suite, the two merging galaxies are only slightly affected during the first pericentric passage, with the inner shells ($r \lesssim 1$~kpc) having an almost constant $l$ and the outer shells ($1 \lesssim r \lesssim 3$~kpc) losing a fraction of their specific angular momentum, both for the gas and (slightly more prominently) for the stars. As expected, the effect is larger in the secondary galaxy, but is somewhat visible also in the primary galaxy, especially in the major mergers (1:1 and 1:2; see Fig.~\ref{angmomflips:fig:m2_hr_gf0_3_BHeff0_001_phi000000_angular_momentum_primary_allgas_allstars_3kpc}). The only exception is the retrograde galaxy in the 1:2 coplanar, prograde--retrograde and retrograde--prograde mergers, which is much less affected by the first pericentric passage than its prograde companion (and than its prograde counterpart in the 1:2 coplanar, prograde--prograde merger), a well-known result in the literature \citep[e.g.][]{Holmberg_1941,Twomre}: stars in a prograde disc are in quasi-resonance with the other galaxy and are either always pulled inwards or outwards, thus creating elongated structures (tails) and changing the angular momentum of the shells; stars in a retrograde disc are instead pulled both inwards and outwards, and therefore do not move much relatively to their own galactic centre \citep[e.g.][]{DOnghia_et_al_2009,DOnghia_et_al_2010}.

{\it Merger stages:} at around the time of the second pericentric passage of {\it all coplanar, prograde--prograde mergers} in the suite, all the gas within the central 3~kpc of the secondary galaxy abruptly changes the polar angle of its angular momentum vector from $\sim$0 to $\sim$$\pi$~radians, which is indicative of ram-pressure shocks significantly altering the dynamics of the gas. This phenomenon is found to be independent of mass ratio (see, e.g., Fig.~\ref{angmomflips:fig:m2_hr_gf0_3_BHeff0_001_phi000000_angular_momentum_secondary_allgas_allstars_3kpc} for a 1:2 case), gas fraction, and BH-physics implementation (see the online-only material). Remarkably, the flip occurs for a large variety of impact parameters, the distance at the second pericentre for these encounters varying from less than 0.1~kpc for the 1:1 merger, to 0.1--0.3~kpc for the 1:2 mergers, to 1.3, 2.5, and 4.3~kpc for the 1:4, 1:6, and 1:10 merger, respectively. The polar angle then remains at $\sim$$\pi$~radians for a significant portion of the merger stage, before the gas counter-flips (with the exception of the 1:1 merger, as described in Section~\ref{angmomflips:sec:1to1coppropro}). However, the time at which the counter-flip occurs varies from merger to merger: the secondary-gas disc changes (again) its rotation at the end of the merger stage in the 1:4, 1:6, and 1:10 mergers, whereas in the 1:2 mergers it does so at around the third pericentric passage, at approximately half the merger stage. This is likely due to a combination of the late pericentric distances (which vary widely from merger to merger) and of how compact the secondary-gas disc is when it encounters the primary-gas disc during these late pericentric encounters.

The non-coplanar, prograde--prograde mergers in the suite behave differently. In the {\it coplanar, prograde--retrograde and retrograde--prograde mergers}, there is no flip, as defined in this paper. The polar angle of the angular momentum of the gas in the secondary galaxy eventually changes from $\sim$0 to $\sim$$\pi$~radians, but it does not do so at around the second pericentric passage. Instead, this happens more gradually and gets completed towards the end of the merger stage, when the two gas discs overlap completely (and indefinitely) and, being the gas dissipative, two counter-rotating gas discs cannot co-exist. The (secondary) stellar flip that we observe around the third pericentric passage is simply due to the fact that the two galaxies are very close and the stars of the primary galaxy dominate. In the {\it inclined-primary mergers} in the suite, also there is no gas flip (as defined in this work), due to the fact that only a limited fraction of the gas in each disc has a direct hydrodynamic interaction with the other, therefore decreasing the effect of the shocks. They show, however, peculiar behaviours that could explain the existence of XRG-like systems (see Section~\ref{angmomflips:sec:1to2incprim}).

{\it Late stages:} in {\it almost all} the systems in the suite, the magnitude of the specific angular momentum of the gas component of the merger remnant is basically the same as that of the primary galaxy during the stochastic stage. This is especially true in the minor mergers (1:4, 1:6, and 1:10). In the 1:2 mergers there is a small difference, due to the fact that the secondary galaxy is not too small compared to the primary. Once again, the only exception is the 1:1 merger (Fig.~\ref{angmomflips:fig:m1_hr_gf0_3_BHeff0_001_phi000000_angular_momentum_secondary_allgas_allstars_3kpc}), where the specific angular momentum of the gas shells with $r \gtrsim 1$~kpc is too chaotic to let us make any comparison (see Section~\ref{angmomflips:sec:1to1coppropro}). On the other hand, the stellar component of the remnant has a lower specific angular momentum than that of the primary galaxy, due to the formation of a large bulge from the partial destruction of the original galactic discs. The difference is minimal in the 1:10 merger (where the primary galactic disc is almost unaffected by the encounter) and, expectedly, increases with the merger mass ratio (compare, e.g., Fig.~\ref{angmomflips:fig:m4_hr_gf0_3_BHeff0_001_phi000000_angular_momentum_primary_allgas_allstars_3kpc} with Fig.~\ref{angmomflips:fig:m2_hr_gf0_3_BHeff0_001_phi000000_angular_momentum_primary_allgas_allstars_3kpc}).

The behaviour described for the default merger about the stellar flip in the inner region ($r \lesssim 0.1$~kpc) of the secondary galaxy (towards the end of the merger stage) and of the merger remnant is {\it not} universal in our suite: it happens in the 1:4 merger (Figs~\ref{angmomflips:fig:m4_hr_gf0_3_BHeff0_001_phi000000_angular_momentum_secondary_allgas_allstars_3kpc} and \ref{angmomflips:fig:m4_hr_gf0_3_BHeff0_001_phi000000_angular_momentum_primary_allgas_allstars_3kpc}), in the 1:2 merger with high BH feedback efficiency, in the 1:2 merger with 60 per cent gas fraction and no BH accretion (in this case the affected region goes out to $\sim$0.3~kpc) and, to some extent, in the 1:6 merger (see the online-only material). Another stellar flip, of a different kind, is observed in the 1:1 merger (see Section~\ref{angmomflips:sec:1to1coppropro}). We will discuss further these stellar flips in Section~\ref{angmomflips:sec:Discussion_and_conclusions}, in the context of KDCs.

\subsubsection{Gas inflows}\label{angmomflips:sec:Mass_inflows}

The second pericentric passage is the time at which the loss of angular momentum is most substantial. We quantify the related gas inflows by calculating how much gas enters a sphere of 200~pc (10 times the gravitational softening of the gas) around the centre of each galaxy.

\begin{figure}
\centering
\vspace{2.5pt}
\includegraphics[trim = 27mm 15mm 6mm 7mm, clip, width=1.20\columnwidth,angle=0]{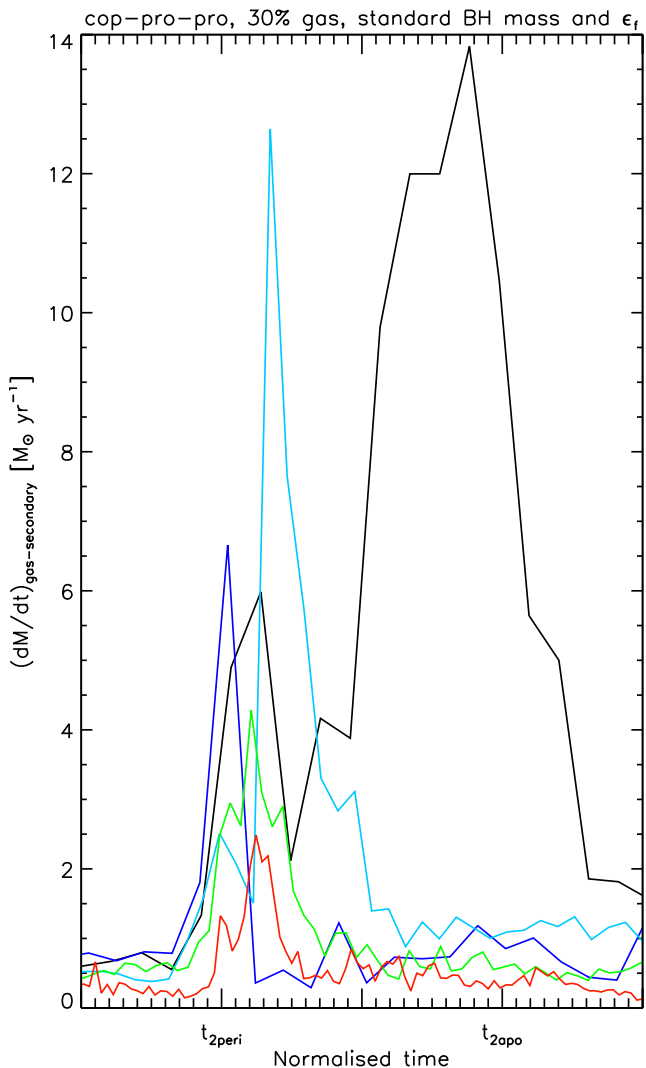}
\vspace{-5pt}
\caption[]{Gas inflow at 200~pc around the centre of the secondary galaxy, for all coplanar, prograde--prograde mergers with 30 per cent gas fraction and standard BH feedback efficiency (1:1: black; 1:2: blue; 1:4: cyan; 1:6: green; and 1:10: red), for a time range that includes the second pericentric and apocentric passages. The time axis has been normalised to allow for an easier comparison.}
\label{angmomflips:fig:Mdot_comparison_allcoplanarlowgasfraction_onepanel_for_flip_sec}
\end{figure}

In Fig.~\ref{angmomflips:fig:Mdot_comparison_allcoplanarlowgasfraction_onepanel_for_flip_sec}, we show the computed mass inflow for the secondary galaxy in each of the five coplanar, prograde--prograde mergers with 30 per cent gas fraction and standard BH mass and feedback efficiency (where the only parameter we vary is the initial mass ratio), for a time range that includes the second pericentric ($t_{\rm 2peri}$) and apocentric ($t_{\rm 2apo}$) passages. Since the orbital history varies depending on the mass ratio, because of the vastly different dynamical friction time-scales \citep[see][]{Capelo_et_al_2015}, we normalised the time axis\footnote{$t_{\rm new} = [t_{\rm old} - (t_{\rm 2peri} + t_{\rm 2apo})/2]/(t_{\rm 2apo} - t_{\rm 2peri}).$} to allow for an easier comparison. In {\it all} the mergers in the figure, there is a substantial increase in $\dot{M}_{\rm gas}$ at around the second pericentric passage, from under 1~M$_{\odot}$~yr$^{-1}$ to several (reaching peaks in the range $\sim$2--13~M$_{\odot}$~yr$^{-1}$, depending on the encounter). After the most violent moment of the encounter ($t_{\rm 2peri}$), the mass inflow goes back to the stochastic levels ($\lesssim$1~M$_{\odot}$~yr$^{-1}$), except for the 1:1 merger, which is a particular case, as explained in Section~\ref{angmomflips:sec:1to1coppropro}. The primary galaxy (not shown) is less affected by the merger, as expected, since the smaller (secondary) companion is usually non large enough to affect its internal structure (again, with the obvious exception of the 1:1 merger, where the two galaxies have equal mass).

\subsection{The 1:1 coplanar, prograde--prograde merger}\label{angmomflips:sec:1to1coppropro}

In this section, we describe the 1:1 coplanar, prograde--prograde merger. We picked this merger because, after the second pericentric passage, it behaves quite differently from all the other coplanar, prograde--prograde mergers in the suite (see Section~\ref{angmomflips:sec:The_other_mergers}).

\begin{figure}
\centering
\vspace{2.5pt}
\includegraphics[trim = 11mm 14mm 0mm 7mm, clip, width=1.20\columnwidth,angle=0]{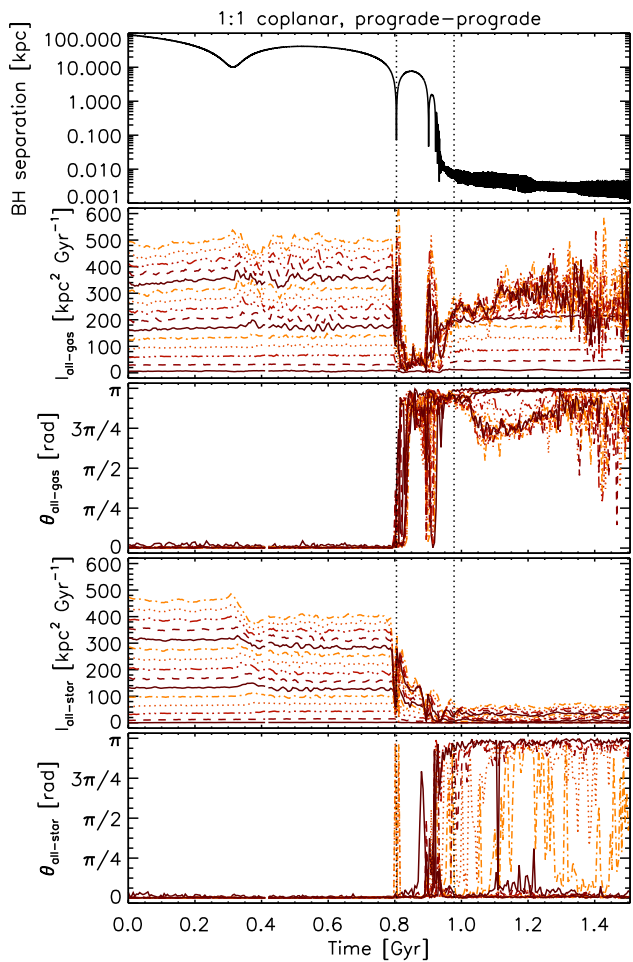}
\vspace{-5pt}
\caption[]{Temporal evolution of the specific angular momentum for one of the galaxies in the 1:1 coplanar, prograde--prograde merger. Same as Fig.~\ref{angmomflips:fig:m4_hr_gf0_3_BHeff0_001_phi000000_angular_momentum_secondary_allgas_allstars_3kpc}.}
\label{angmomflips:fig:m1_hr_gf0_3_BHeff0_001_phi000000_angular_momentum_secondary_allgas_allstars_3kpc}
\end{figure}

Fig.~\ref{angmomflips:fig:m1_hr_gf0_3_BHeff0_001_phi000000_angular_momentum_secondary_allgas_allstars_3kpc} shows the evolution of one of the galaxies (the evolution of the other galaxy, not shown in the paper, is almost identical) of this major merger. At the beginning, the behaviour is in some ways similar to that of the default merger: the gas flips at around the second pericentric passage, while its angular momentum drops almost to zero. However, after the end of the merger stage, we note that the gas beyond 1~kpc has not returned to a ``stable'' configuration like in the case of the default merger (only the inner-kpc $l$ curves return to being ``flat'' during the remnant stage) and, moreover, the gas has not counter-flipped. Most importantly, stars also behave quite differently. The stellar specific angular momentum almost drops to zero towards the end of the merger stage and stays extremely low during the entire remnant stage. Additionally, most of the stars -- in particular, almost all the stars in the inner kpc -- have also flipped, although not at the beginning of the merger stage (as the gas did), but towards its end.

This can be explained by the fact that these two galaxies ``hit'' each other in an almost radial orbit (the second pericentric distance is less than 100~pc, the smallest in our entire suite) and have equal mass, causing the almost complete disruption of the two merging systems. When the merger remnant forms, stars have lost any coherent rotation and therefore have an almost zero net angular momentum\footnote{Thanks to the inflow of (dissipative) gas from the outer regions, the inner shells retain some net angular momentum. This does not happen to the stellar (dissipationless) component.}. The end result is a merger remnant in which the gas and most of the inner-kpc stars co-rotate in a retrograde manner, and a significant fraction of the outer stars ($1 \lesssim r \lesssim 3$~kpc) are counter-rotating with respect to both the gas and the inner-kpc stars, therefore keeping their initial prograde rotation. Note, however, that the specific angular momentum of all these stars, especially those in the inner kpc, is very low (due to the disruption of the two galactic discs). In Section~\ref{angmomflips:sec:Discussion_and_conclusions} we will discuss this phenomenon in relation to KDCs.

\begin{figure}
\centering
\vspace{2.5pt}
\includegraphics[trim = 11mm 14mm 0mm 7mm, clip, width=1.20\columnwidth,angle=0]{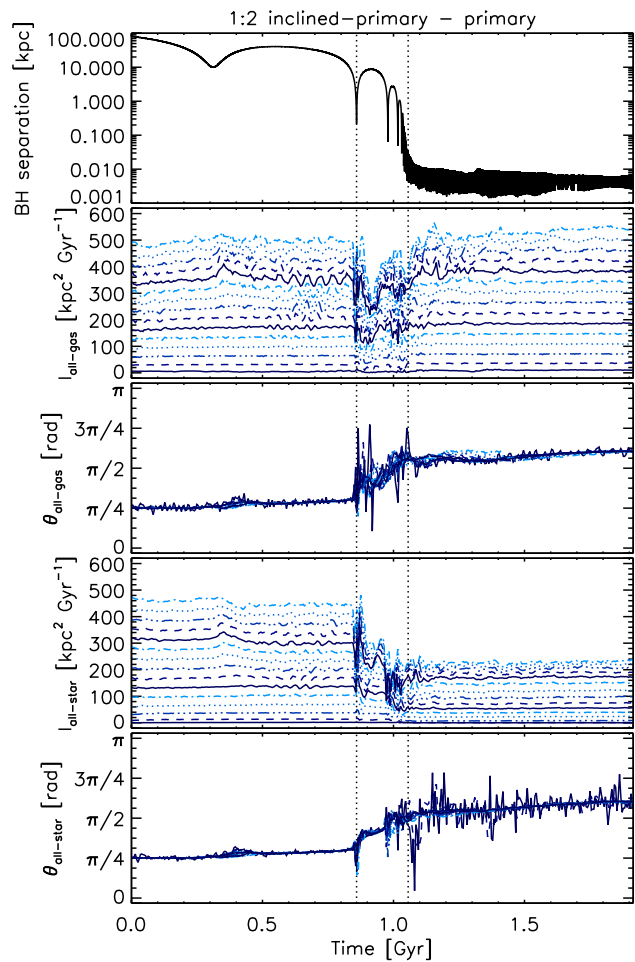}
\vspace{-5pt}
\caption[]{Temporal evolution of the specific angular momentum for the primary galaxy in the 1:2 inclined-primary merger. Same as Fig.~\ref{angmomflips:fig:m4_hr_gf0_3_BHeff0_001_phi000000_angular_momentum_secondary_allgas_allstars_3kpc}.}
\label{angmomflips:fig:m2_hr_gf0_3_BHeff0_001_phi045000_angular_momentum_primary_allgas_allstars_3kpc}
\end{figure}


\subsection{The 1:2 inclined-primary merger}\label{angmomflips:sec:1to2incprim}

In this section, we describe the 1:2 inclined-primary merger, both as an example of a non-coplanar merger and also because of its possible connection to XRG-like systems. Fig.~\ref{angmomflips:fig:m2_hr_gf0_3_BHeff0_001_phi045000_angular_momentum_primary_allgas_allstars_3kpc} shows the evolution of the primary galaxy (the evolution of the secondary galaxy is shown in the online-only material). We focus on the primary galaxy because, contrary to what happens in most of the mergers in our suite, the larger galaxy is as heavily affected as the smaller one. Moreover, both the gas and the stars respond to the merger. The evolution of the magnitude of the specific angular momentum of both stars and gas, in both the primary and secondary galaxies, is similar to that of the coplanar, prograde--prograde mergers (see Figs~\ref{angmomflips:fig:m2_hr_gf0_3_BHeff0_001_phi000000_angular_momentum_secondary_allgas_allstars_3kpc}--\ref{angmomflips:fig:m2_hr_gf0_3_BHeff0_001_phi000000_angular_momentum_primary_allgas_allstars_3kpc}). However, the polar angle of both baryonic species (gradually) changes during the merger stage, from the initial $\pi/4$ to $\sim$$5\pi/8$~radians (at the same time, the change in the secondary galaxy is from the initial 0 to $\sim$$5\pi/8$~radians; see the online-only material). During the remnant stage, all shells within 3~kpc share the same polar angle (for both stars and gas, in both galaxies), except for the very inner region ($r \lesssim 300$~pc), where we observe some minor fluctuations.

We argue, however, that this is not necessarily a hydrodynamically driven phenomenon (as opposed to that described for the default merger in Section ~\ref{angmomflips:sec:1to4coppropro}, for example), but it is likely due to a purely gravitational interaction, since both stars and gas are affected.

The implications of such change in the polar angle of the angular momentum of the gas discs will be further discussed in Section~\ref{angmomflips:sec:Discussion_and_conclusions}, where we briefly speculate that phenomena such as this could be at the origin of systems similar to some of the observed XRGs.


\section{Discussion and Conclusions}\label{angmomflips:sec:Discussion_and_conclusions}

In this study we highlighted the importance of hydrodynamical effects during close pericentric passages of galaxy encounters in inducing gas inflows and angular momentum flips. The hydrodynamic shocks developing in the ISM at the contact surface of the two merging structures modify dramatically the angular momentum budget of the gas distribution, effectively decoupling the gaseous component from the stellar one. All the interactions we studied trigger large gas inflows toward the central regions either of the secondary or of both the hosts. Such inflows result in intense bursts of central star formation and could, if a fraction of the infalling gas proceeds unimpeded to sub-pc (unresolved) scales, fuel single and dual AGN activity (\citealt{Capelo_et_al_2015}, \citeyear{Capelo_et_al_2016}). These gas inflows also result in the formation of dynamically decoupled stellar structures, mostly dominated by young stars.

There is an important difference between the dynamical properties of the shock-driven gas inflows and the inflows associated to the tidal triggering of stellar non-axisymmetric structures discussed in Section~\ref{angmomflips:sec:Introduction}: as the former process results in a sudden decoupling of the gas dynamics from the stellar one, the angular momentum distribution of the inflowing gas immediately after the pericentric passage depends on the parameters of the galactic impact. For some configurations, the shocks can populate a region in the parameter space of extremely low angular momentum. Such region could be unaccessible to the latter processes, for example if the matter distribution of the two galaxies supports some orbital resonances associated to the triggered non-axisymmetric structure, at which the gas inflow slows down or stops (see e.g. \citealt{Kormendy_2013} for a recent discussion). For this reason, the newly proposed mechanism is particularly interesting as an efficient driver of very compact starbursts (within less than $\sim$100~pc) and BH accretion well before the two galaxies complete their merger.

It is remarkable that the gas flip caused by these ram-pressure shocks is an almost universal occurrence in our coplanar, prograde--prograde mergers, regardless of the wide variety of impact parameters (i.e. second pericentric distances: 0.1--4.3~kpc) we simulated. This is in contrast to the fact that such phenomenon has not been reported before, especially considering the vast number of coplanar encounters studied in the past three decades. We note that all of our encounters share the same first pericentric distance ($\sim$10~kpc, i.e 20 per cent of the virial radius of the primary galaxy), equal to $\sim$10--20 times the scale radii of the two galactic discs. This value was chosen in accordance to what found in cosmological simulations \citep{Khochfar2006}, which show that 85 per cent of merging halo orbits have a first pericentric distance greater than 10 per cent of the virial radius of the primary galaxy. In comparison to this, simulations of merging galaxies in the past \citep[e.g.][]{PDiMatteo_et_al_2007,Cox2008,PDiMatteo_et_al_2008,Younger2008,Johansson2009} have usually adopted a smaller `first pericentric distance'-to-`disc scale radius' ratio (one of the reasons being that a smaller first pericentric distance makes the merging faster and therefore computationally less expensive to simulate). One consequence of this choice, however, is that the two galaxies would be more heavily affected by their first encounter, thereby changing their own internal structure also in the more central regions\footnote{We note, in passing, that this is the likely reason why the central star formation and AGN activity of our galaxies after the first pericentric passage \citep{Capelo_et_al_2015} change less than in other studies \citep[e.g.][]{Younger2008,Johansson2009}.}. By adopting a larger (and more realistic) first impact parameter, we had to simulate for a relatively longer time (i.e. for $\sim$1~Gyr before the second pericentre), but also managed to recover this new and interesting result.

We argue that a more thorough study, involving a variety of initial global orbital parameters (e.g. type of orbit; first pericentric distance), is needed in order to better quantify the effect of the orbits on the phenomena described in this paper. Moreover, varying also the internal orbital parameters (i.e. the initial direction of the galactic angular momentum) would allow us to determine for which range of angles the flip occurs. This last point is particularly important, since in inclined mergers ($i$) the hydrodynamic interaction between the two gas discs is limited, therefore decreasing the effect of the shocks, and ($ii$) the efficiency of merger-induced torques is lower than that in coplanar encounters \citep[e.g.][]{Cox2008,Capelo_et_al_2015}. For such a study, given the high number of encounters needed and in order to have relatively cheap computations, one could in principle perform simulations without some of the sub-grid models which are not strictly necessary to study the fate of the gas (e.g. star formation, stellar feedback, and BHs). We stress, however, that a high resolution is still needed to reliably resolve the shock between the two discs and to verify the aftermath of the shock itself, in particular when one is interested in nuclear gas inflows.

The formation of long-lived (up to Gyr time-scales) dynamically decoupled gas and newly-formed star structures, clearly visible in the default merger (Figs~\ref{angmomflips:fig:m4_hr_gf0_3_BHeff0_001_phi000000_angular_momentum_secondary_allgas_allstars_3kpc} and \ref{angmomflips:fig:m4_hr_gf0_3_BHeff0_001_phi000000_angular_momentum_primary_allgas_allstars_3kpc}) and in two of the 1:2 coplanar, prograde--prograde mergers (see the online-only material), suggests that the hydrodynamic mechanism presented in this paper could represent an alternative channel\footnote{For the previously proposed mechanisms, see e.g. \citet{Balcells_Quinn_1990,Jesseit_et_al_2007,Bois_et_al_2010,Bois_et_al_2011,Tsatsi_et_al_2015}.} for the formation of KDCs commonly observed in early-type galaxies \citep[e.g.][]{McDermid_et_al_2006}.  KDCs are often distinguished into two categories: {\it large} KDCs, with characteristic scale lengths up to $\sim$1~kpc, formed by old stars (consistent with the age of the overall stellar distribution) and typically hosted by ``slow rotators'' (i.e. galaxies without dynamically significant rotation; see e.g. \citealt{Emsellem_et_al_2007,Cappellari_et_al_2007}), and {\it small} KDCs, hosted in ``fast rotators'', confined into smaller nuclear regions ($\sim$100~pc) and showing hints of recent star formation \citep[e.g.][]{McDermid_et_al_2006}. The hydrodynamical process we described could be particularly related to the formation of more compact KDCs in fast rotators. The dynamically decoupled gas inflow allows for the build up of KDCs with recent star formation. In addition to that, we notice that usually the kinematically misaligned component is limited to the central $\sim$100~pc. A noticeable exception is represented by the 1:1 coplanar, prograde--prograde merger, for which the interaction is sufficiently intense to erase most of the angular momentum of the remnant. In this case, the decoupled gas--stellar structure extends out to $\sim$1~kpc and it is long lived (remaining present till the end of the run), reminiscent of the largest KDCs in slow rotators.

Lastly, we wish to briefly speculate on the intriguing implications of the change in direction of the galactic angular momentum of the merging galaxies. If the infalling gas has a different angular momentum orientation with respect to the original gas distribution in the galactic nuclei, then a chain of successive shocks on smaller and smaller scales (unresolved in this study) could contribute to further remove the gas angular momentum, possibly driving AGN activity \citep[][]{Hobbs_et_al_2011,Carmona_et_al_2014,Carmona_et_al_2015}. If we assume that the gas can proceed unimpeded to sub-pc scales, down to the formation of a circumnuclear disc and further down to that of an accretion disc around the central BH (or BH binary), then the direction of the galactic angular momentum and that of the central BH's spin may be related \citep[e.g.][]{bogdanovic2007,Dotti2010}. The direction of the BH's spin can in turn determine the direction of (possible) jets \citep[e.g.][]{Blandford_Znajek_1977}. If the change in the galactic angular momentum angle translates to a similar change in the BH's spin, then also the direction of the jets is similarly affected. For example, in the inclined mergers, the angular momentum
of the gas can change significantly ($\sim$$\pi/4$ radians for the secondary galaxy in the 1:4 merger; $3\pi/8$ and $5\pi/8$~radians for the primary and secondary galaxy, respectively, in the 1:2 merger; see the online-only material and Fig.~\ref{angmomflips:fig:m2_hr_gf0_3_BHeff0_001_phi045000_angular_momentum_primary_allgas_allstars_3kpc}), resulting in possibly very differently oriented radio jets and thus creating an ``X'' shape similar to that of the XRGs \citep[][]{Leahy_Parma_1992} observed in more massive systems. Of course, further studies with improved resolution (e.g. $<$1~pc) should be pursued, to check that the reported general behaviour of the gas angular momentum holds at smaller scales.

\section*{Acknowledgements}
We thank the reviewer, Joshua~E.~Barnes, for the useful comments that greatly improved this work. We also thank Davide Fiacconi and Alessandro Lupi for a thorough reading of the manuscript. HPC resources supporting this work were provided by the NASA High-End Computing (HEC) Program through the NASA Advanced Supercomputing (NAS) Division at Ames Research Center, and by TGCC, under the allocations 2013-t2013046955 and 2014-x2014046955 made by GENCI. PRC acknowledges support by the Tomalla Foundation.

\scalefont{0.94}
\setlength{\bibhang}{1.6em}
\setlength\labelwidth{0.0em}
\bibliographystyle{mnras}
\bibliography{flip_paper}
\normalsize

\appendix

\section{Angular momentum evolution of select galaxies}\label{angmomflips:sec:Additional_figures}

\begin{figure}
\centering
\vspace{2.5pt}
\includegraphics[trim = 11mm 14mm 30mm 7mm, clip, width=0.8\columnwidth,angle=0]{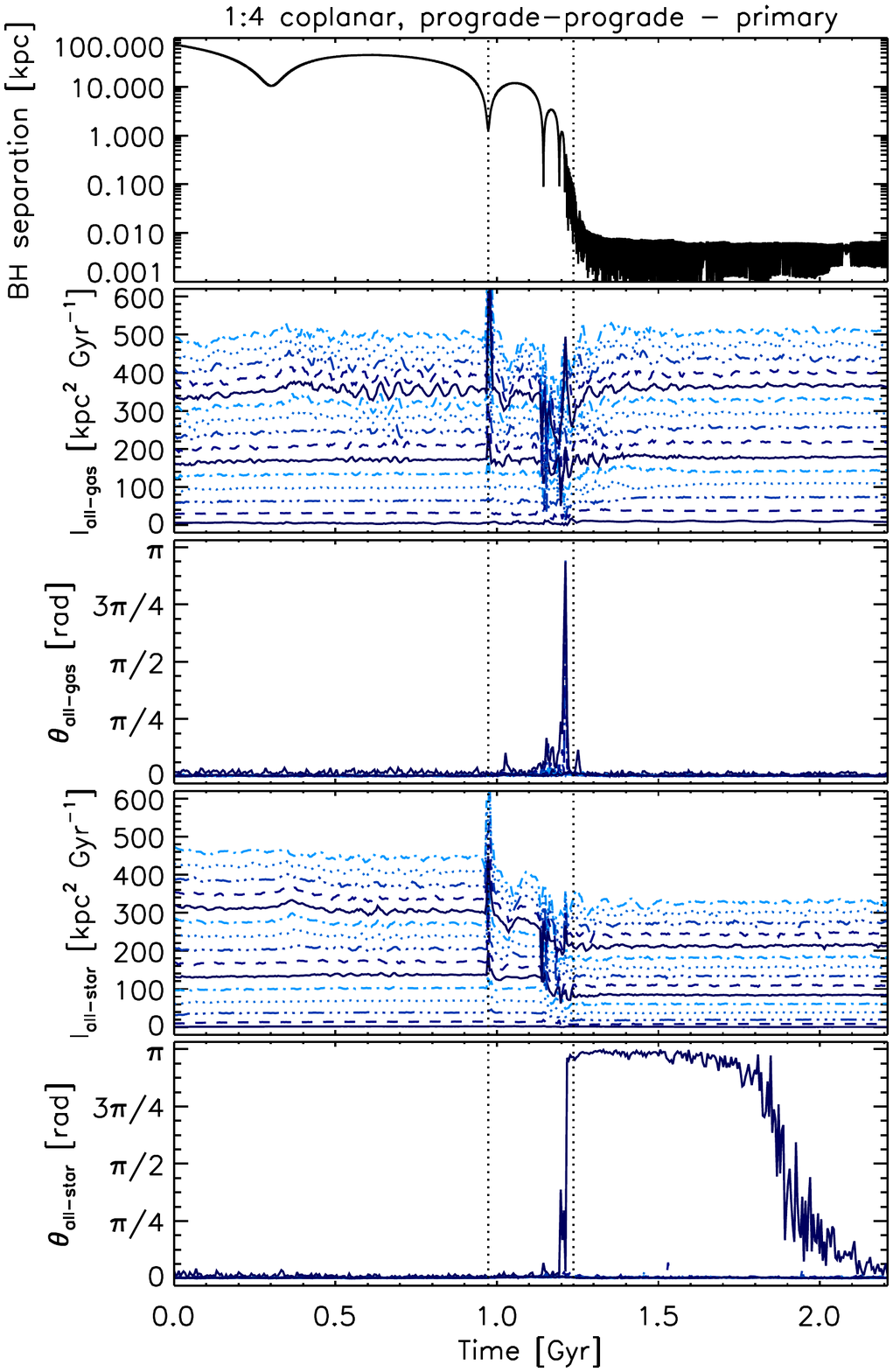}
\vspace{-5pt}
\caption[]{Temporal evolution of the specific angular momentum for the primary galaxy in the 1:4 coplanar, prograde--prograde merger. Same as Fig.~\ref{angmomflips:fig:m4_hr_gf0_3_BHeff0_001_phi000000_angular_momentum_secondary_allgas_allstars_3kpc}.\\}
\label{angmomflips:fig:m4_hr_gf0_3_BHeff0_001_phi000000_angular_momentum_primary_allgas_allstars_3kpc}
\end{figure}

\begin{figure}
\centering
\vspace{2.5pt}
\includegraphics[trim = 11mm 14mm 30mm 7mm, clip, width=0.8\columnwidth,angle=0]{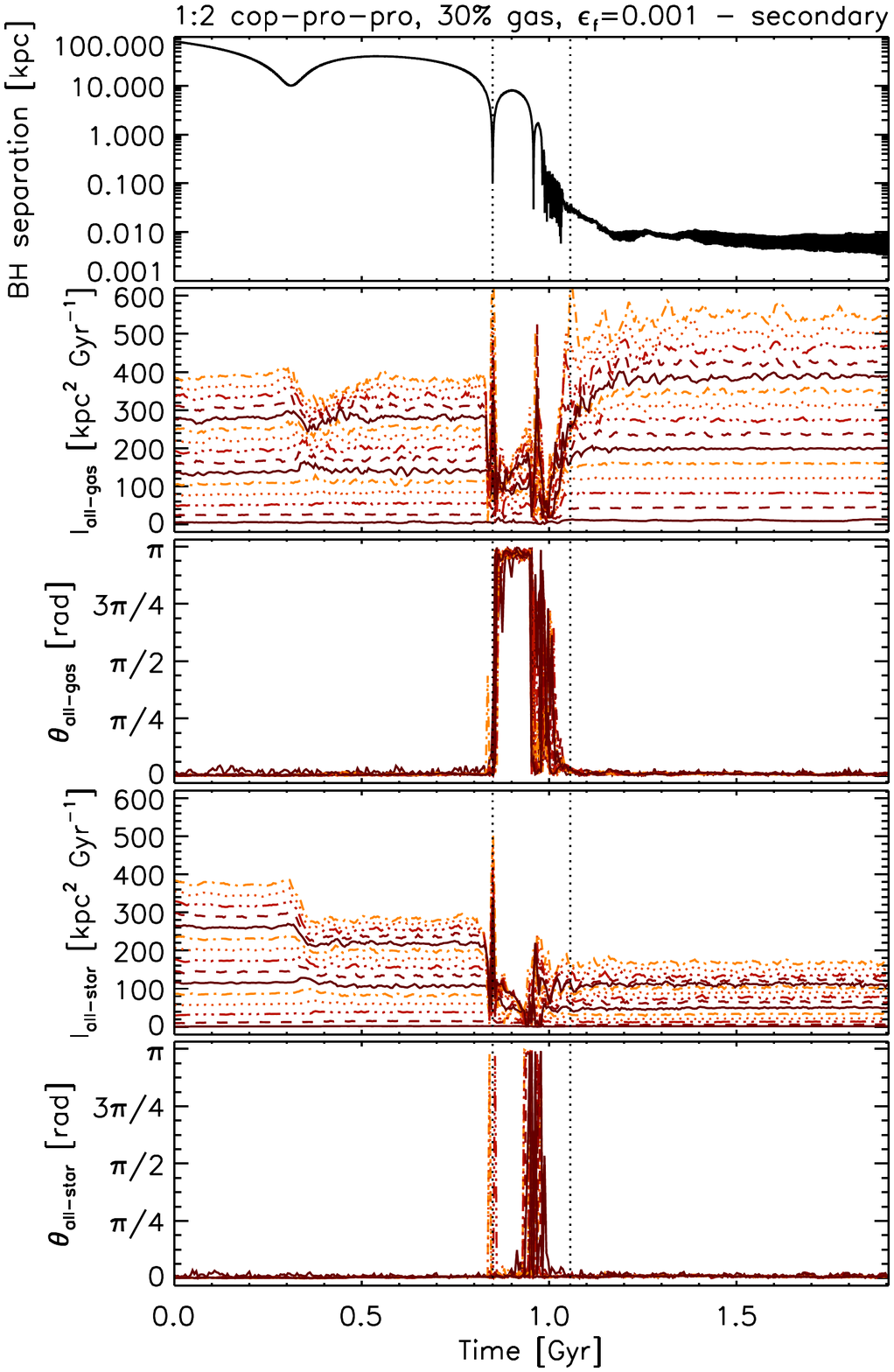}
\vspace{-5pt}
\caption[]{Temporal evolution of the specific angular momentum for the secondary galaxy in the 1:2 coplanar, prograde--prograde merger with 30 per cent gas fraction and standard BH mass and feedback efficiency. Same as Fig.~\ref{angmomflips:fig:m4_hr_gf0_3_BHeff0_001_phi000000_angular_momentum_secondary_allgas_allstars_3kpc}.}
\label{angmomflips:fig:m2_hr_gf0_3_BHeff0_001_phi000000_angular_momentum_secondary_allgas_allstars_3kpc}
\end{figure}

\begin{figure}
\centering
\vspace{2.5pt}
\includegraphics[trim = 11mm 14mm 30mm 7mm, clip, width=0.8\columnwidth,angle=0]{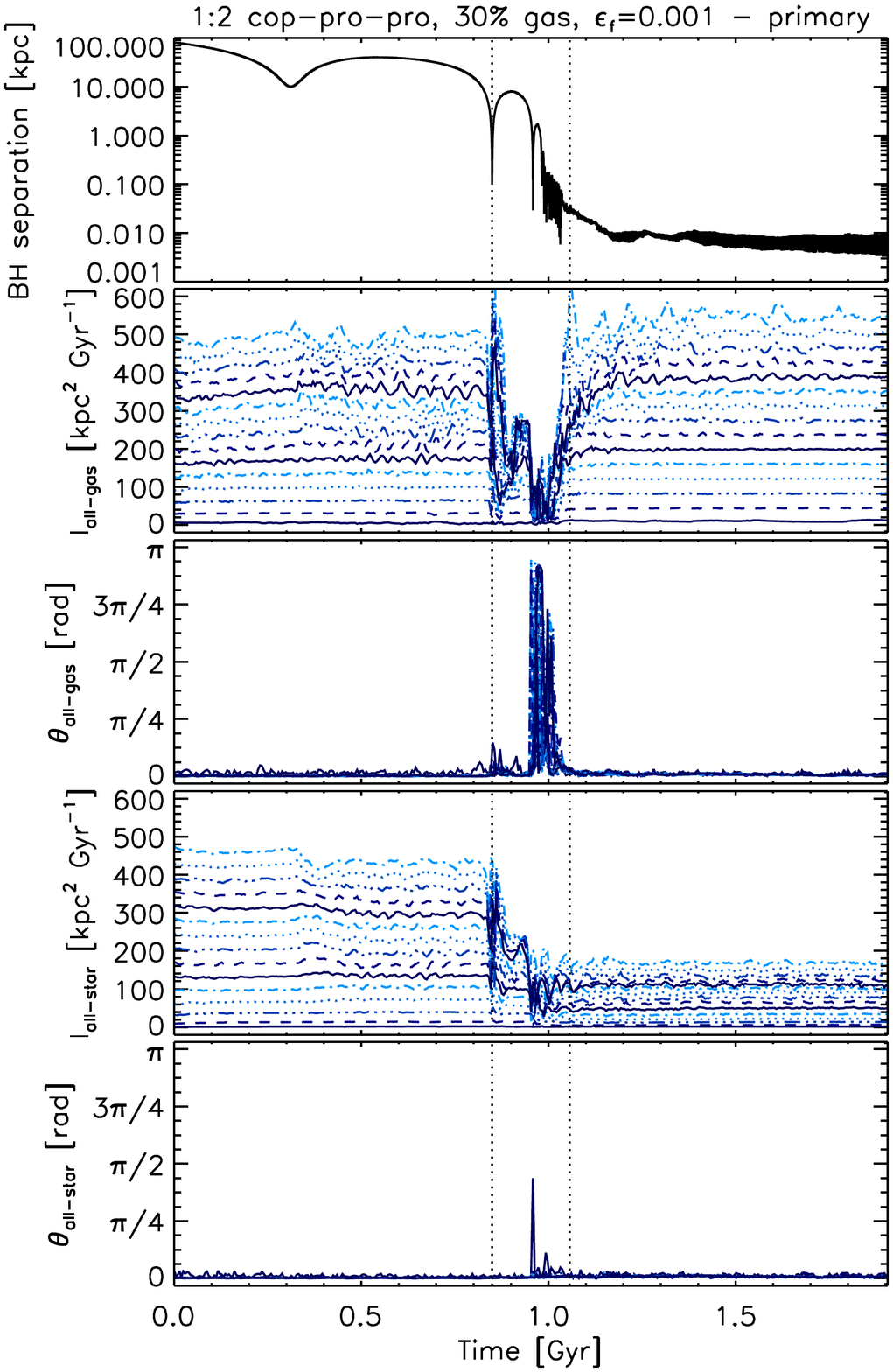}
\vspace{-5pt}
\caption[]{Temporal evolution of the specific angular momentum for the primary galaxy in the 1:2 coplanar, prograde--prograde merger with 30 per cent gas fraction and standard BH mass and feedback efficiency. Same as Fig.~\ref{angmomflips:fig:m4_hr_gf0_3_BHeff0_001_phi000000_angular_momentum_secondary_allgas_allstars_3kpc}.}
\label{angmomflips:fig:m2_hr_gf0_3_BHeff0_001_phi000000_angular_momentum_primary_allgas_allstars_3kpc}
\end{figure}

\clearpage

\section{Online-only supplementary figures}\label{angmomflips:sec:Online_only_material}

\begin{figure}
\centering
\vspace{2.5pt}
\includegraphics[trim = 11mm 14mm 30mm 7mm, clip, width=0.8\columnwidth,angle=0]{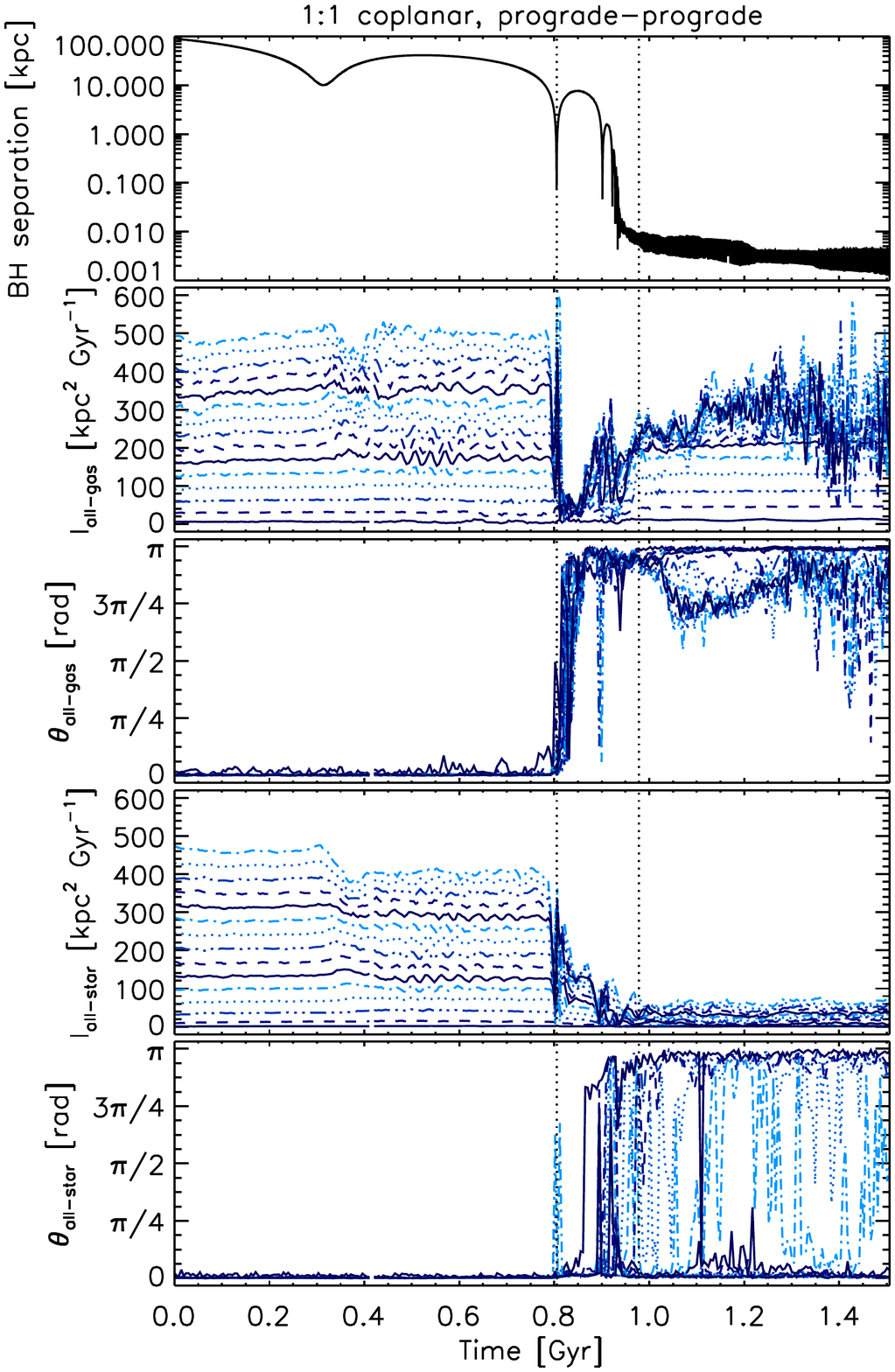}
\vspace{-5pt}
\caption[]{Temporal evolution of the specific angular momentum for one of the galaxies in the 1:1 coplanar, prograde--prograde merger. Same as Fig.~1 in the main text.}
\label{angmomflips:fig:m1_hr_gf0_3_BHeff0_001_phi000000_angular_momentum_primary_allgas_allstars_3kpc}
\end{figure}

\begin{figure}
\centering
\vspace{2.5pt}
\includegraphics[trim = 11mm 14mm 30mm 7mm, clip, width=0.8\columnwidth,angle=0]{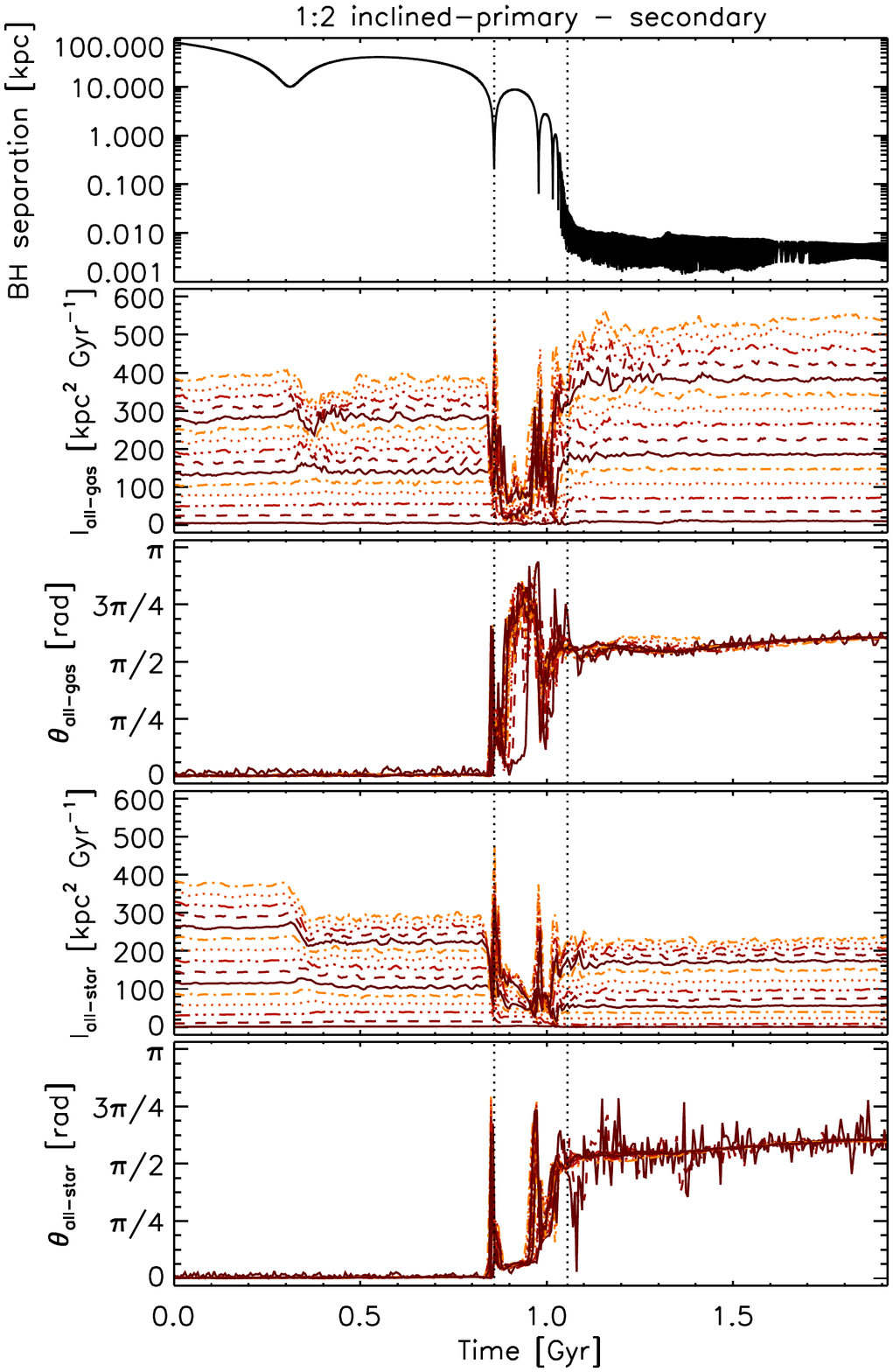}
\vspace{-5pt}
\caption[]{Temporal evolution of the specific angular momentum for the secondary galaxy in the 1:2 inclined-primary merger. Same as Fig.~1 in the main text.}
\label{angmomflips:fig:m2_hr_gf0_3_BHeff0_001_phi045000_angular_momentum_secondary_allgas_allstars_3kpc}
\end{figure}

\begin{figure*}
\centering
\vspace{2.5pt}
\includegraphics[trim = 11mm 14mm 30mm 7mm, clip, width=0.8\columnwidth,angle=0]{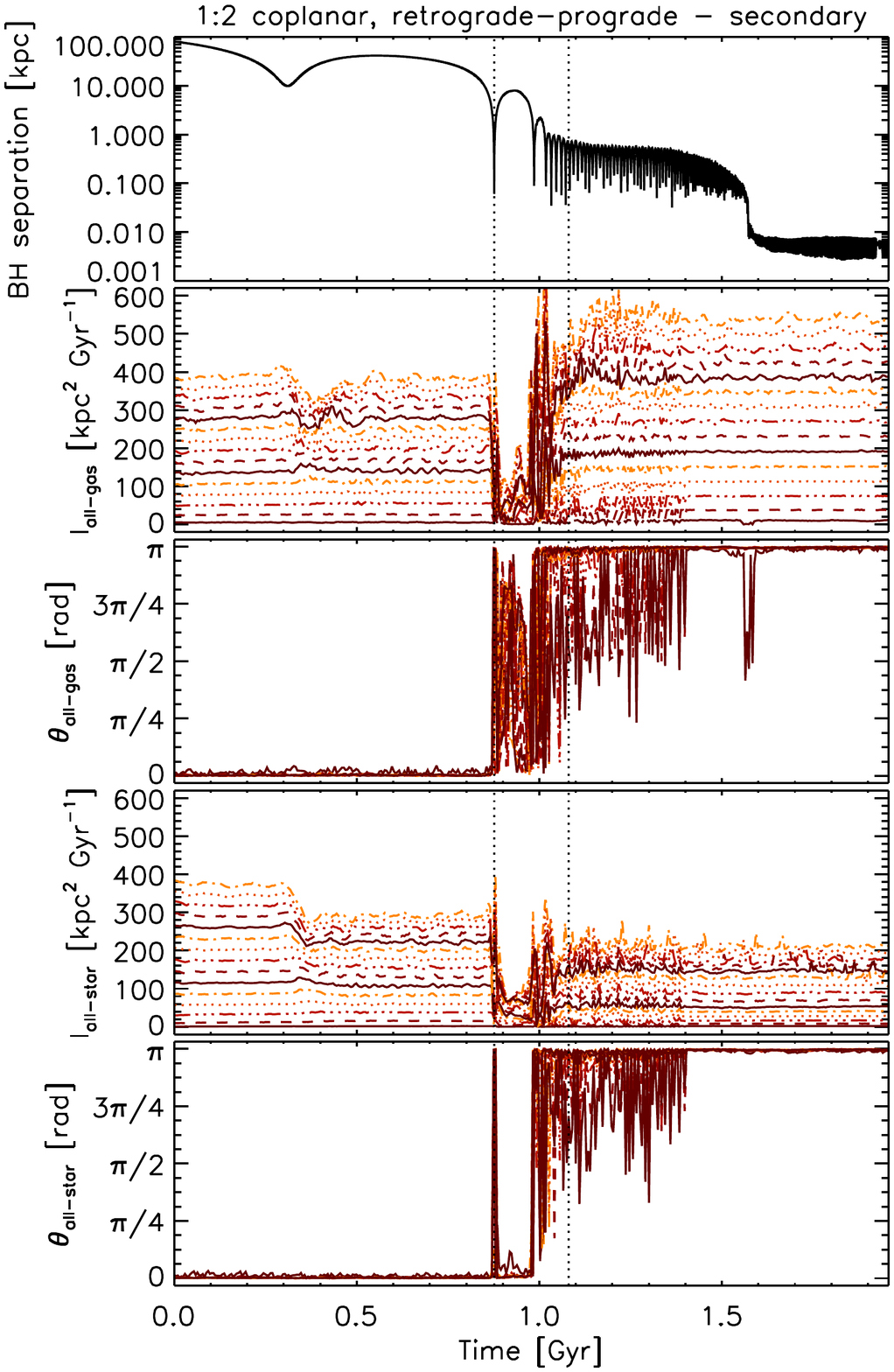}
\hspace{50.0pt}\includegraphics[trim = 11mm 14mm 30mm 7mm, clip, width=0.8\columnwidth,angle=0]{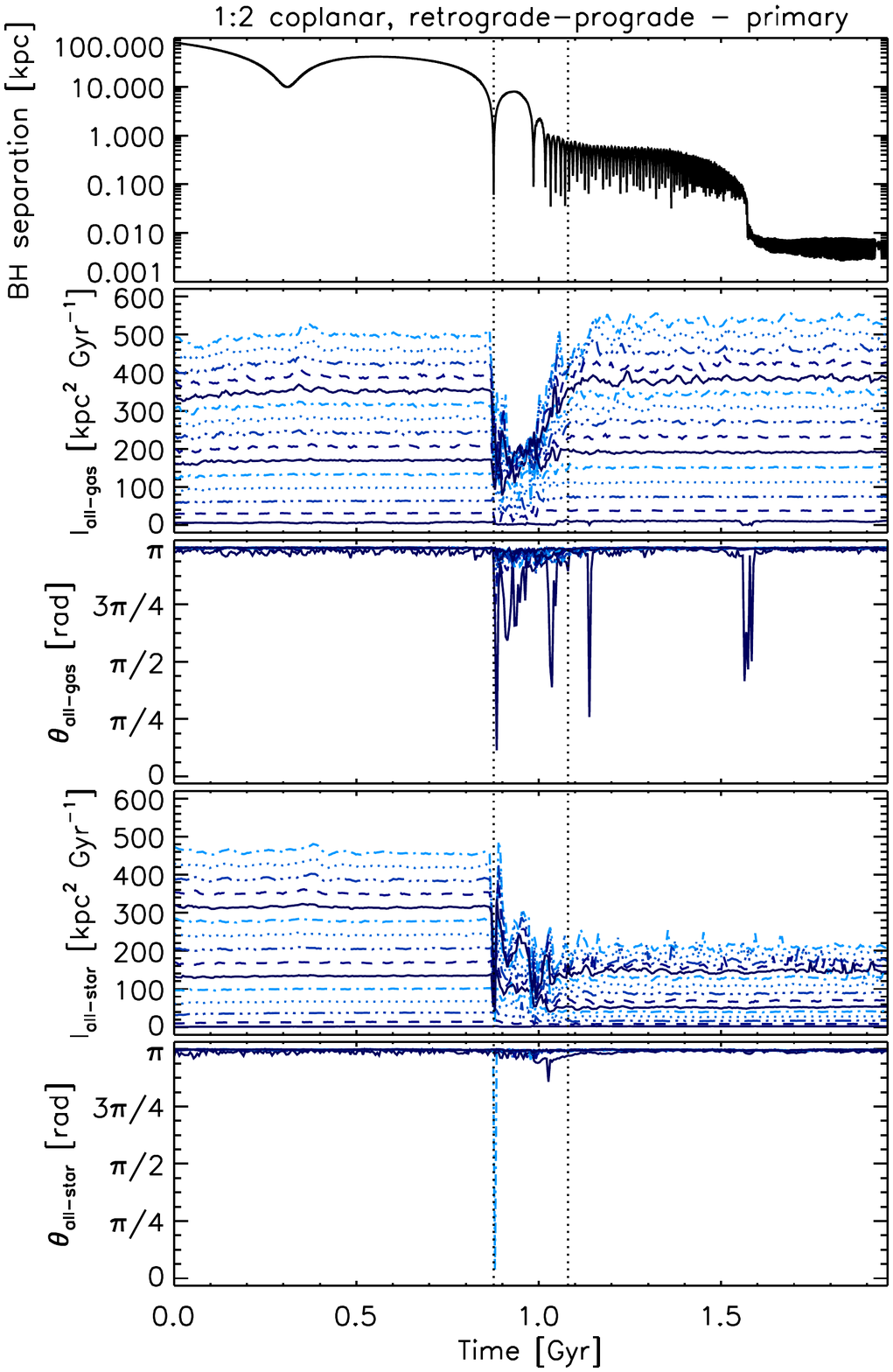}
\vspace{-5pt}
\caption[]{Temporal evolution of the specific angular momentum for the secondary (left) and primary (right) galaxy in the 1:2 coplanar, retrograde--prograde merger. Same as Fig.~1 in the main text.}
\label{angmomflips:fig:m2_hr_gf0_3_BHeff0_001_phi180000_angular_momentum_allgas_allstars_3kpc}
\end{figure*}

\begin{figure*}
\centering
\vspace{2.5pt}
\includegraphics[trim = 11mm 14mm 30mm 7mm, clip, width=0.8\columnwidth,angle=0]{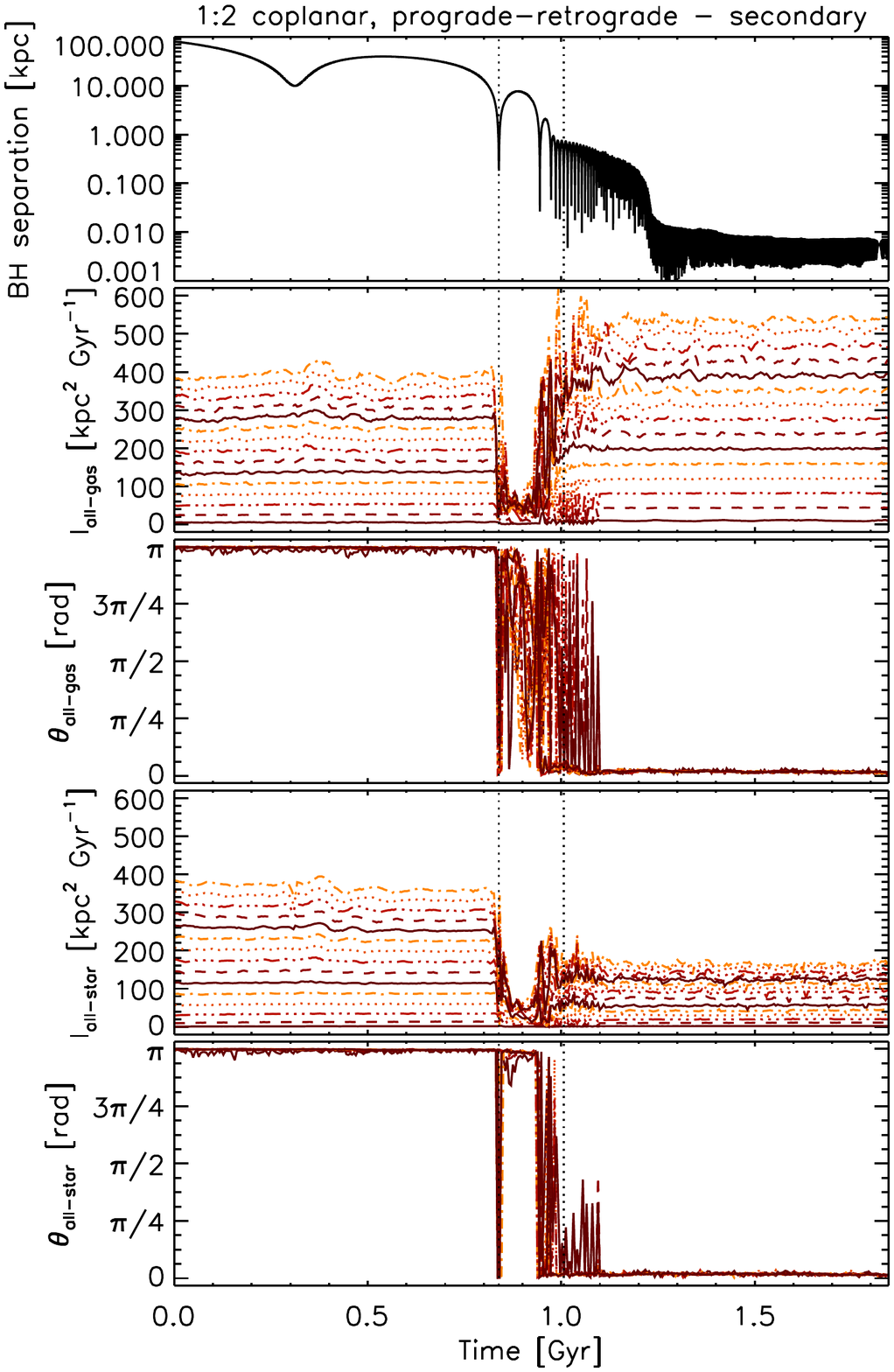}
\hspace{50.0pt}\includegraphics[trim = 11mm 14mm 30mm 7mm, clip, width=0.8\columnwidth,angle=0]{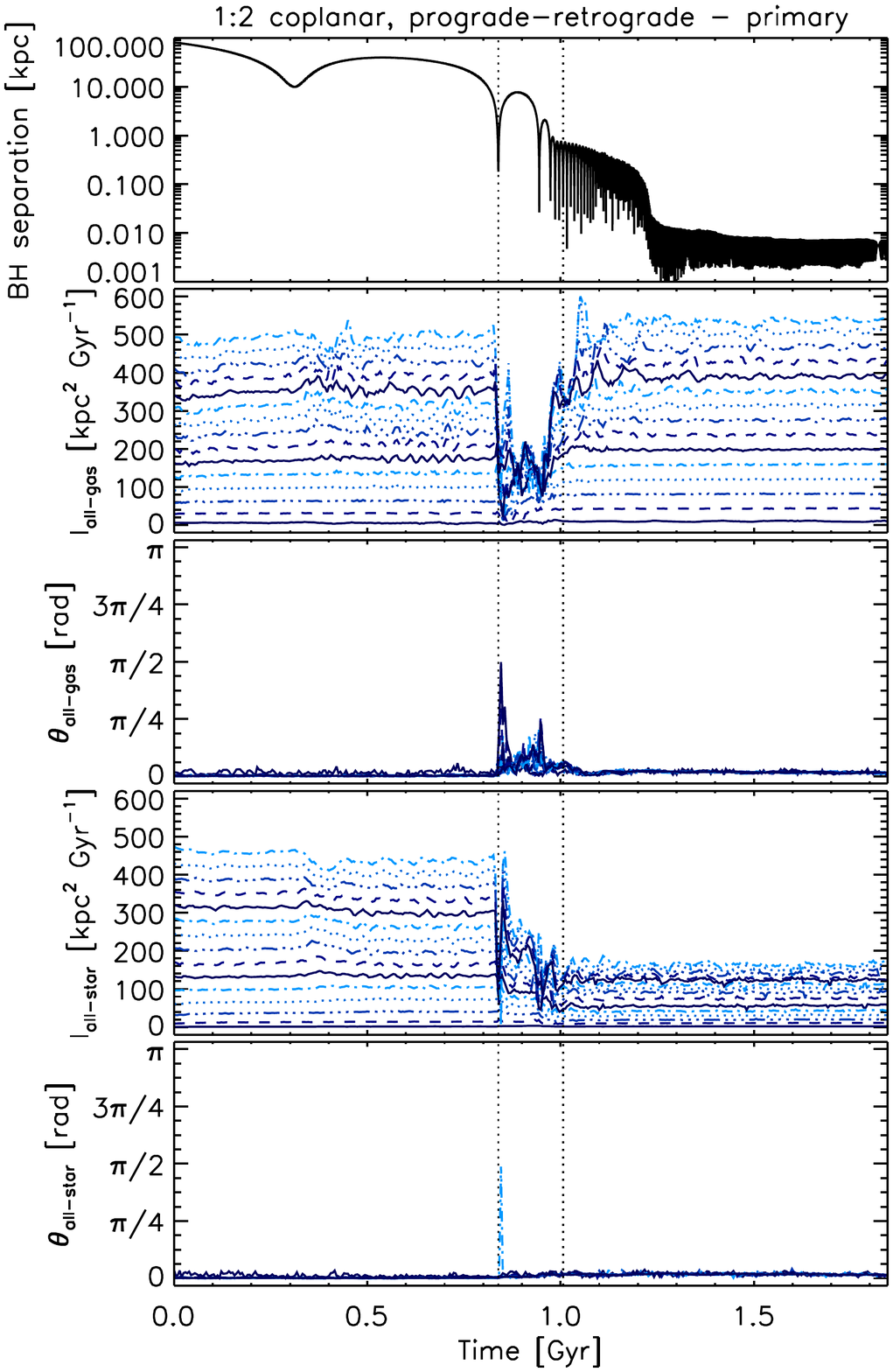}
\vspace{-5pt}
\caption[]{Temporal evolution of the specific angular momentum for the secondary (left) and primary (right) galaxy in the 1:2 coplanar, prograde--retrograde merger. Same as Fig.~1 in the main text.}
\label{angmomflips:fig:m2_hr_gf0_3_BHeff0_001_phi000180_angular_momentum_allgas_allstars_3kpc}
\end{figure*}

\begin{figure*}
\centering
\vspace{2.5pt}
\includegraphics[trim = 11mm 14mm 30mm 7mm, clip, width=0.8\columnwidth,angle=0]{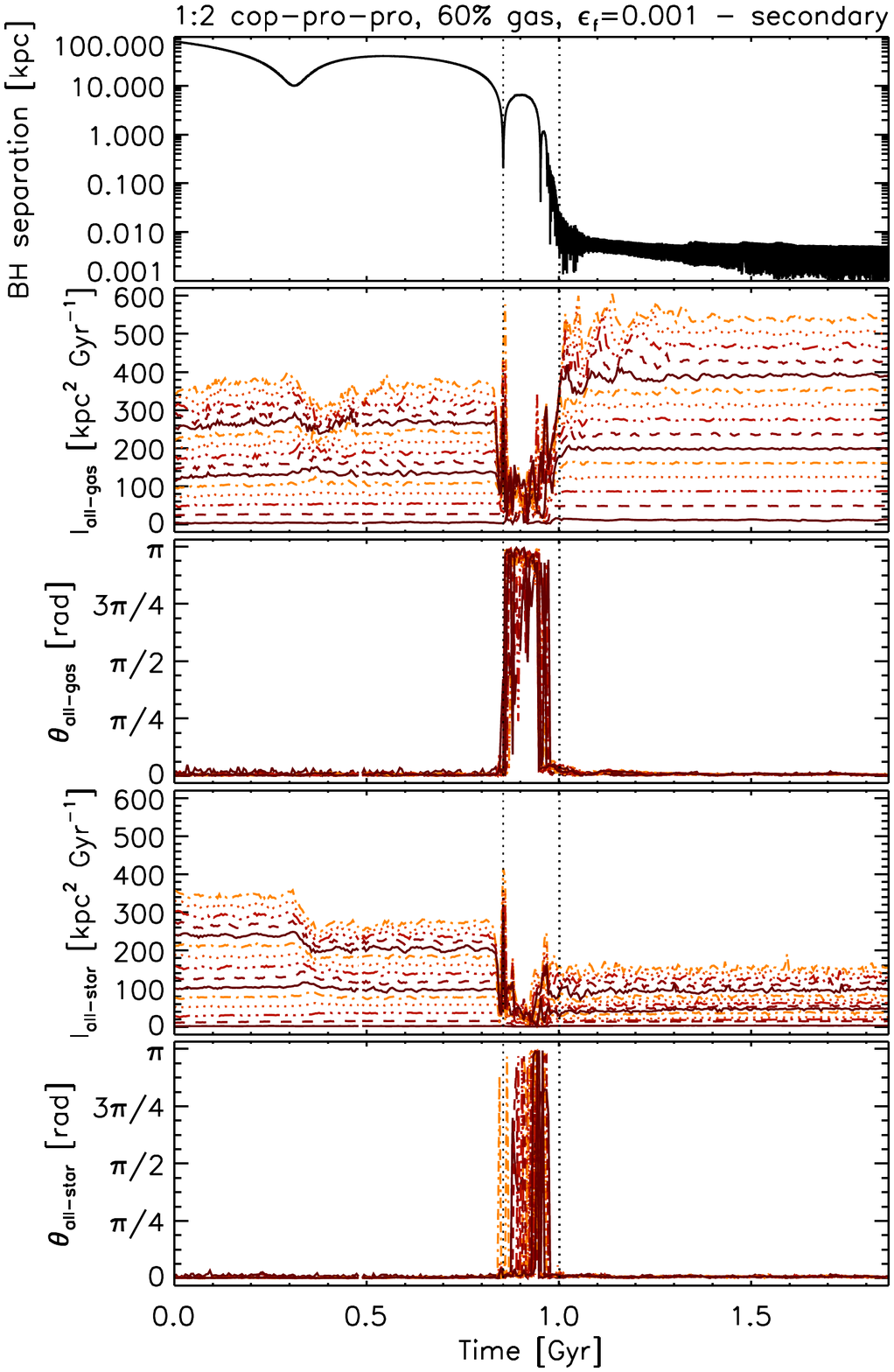}
\hspace{50.0pt}\includegraphics[trim = 11mm 14mm 30mm 7mm, clip, width=0.8\columnwidth,angle=0]{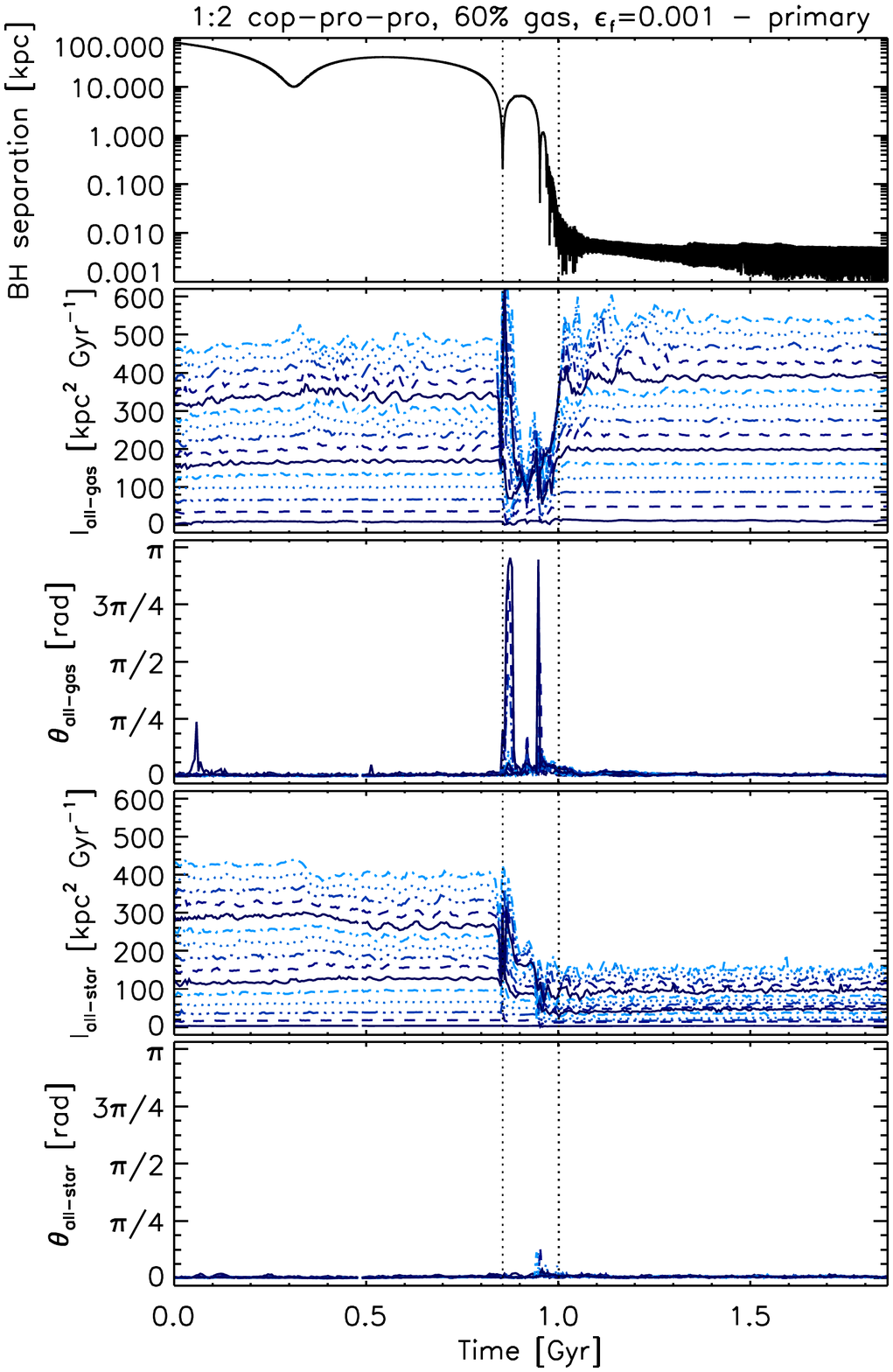}
\vspace{-5pt}
\caption[]{Temporal evolution of the specific angular momentum for the secondary (left) and primary (right) galaxy in the 1:2 coplanar, prograde--prograde merger with 60 per cent gas fraction and standard BH feedback efficiency. Same as Fig.~1 in the main text.}
\label{angmomflips:fig:m2_hr_gf0_6_BHeff0_001_phi000000_angular_momentum_allgas_allstars_3kpc}
\end{figure*}

\begin{figure*}
\centering
\vspace{2.5pt}
\includegraphics[trim = 11mm 14mm 30mm 7mm, clip, width=0.8\columnwidth,angle=0]{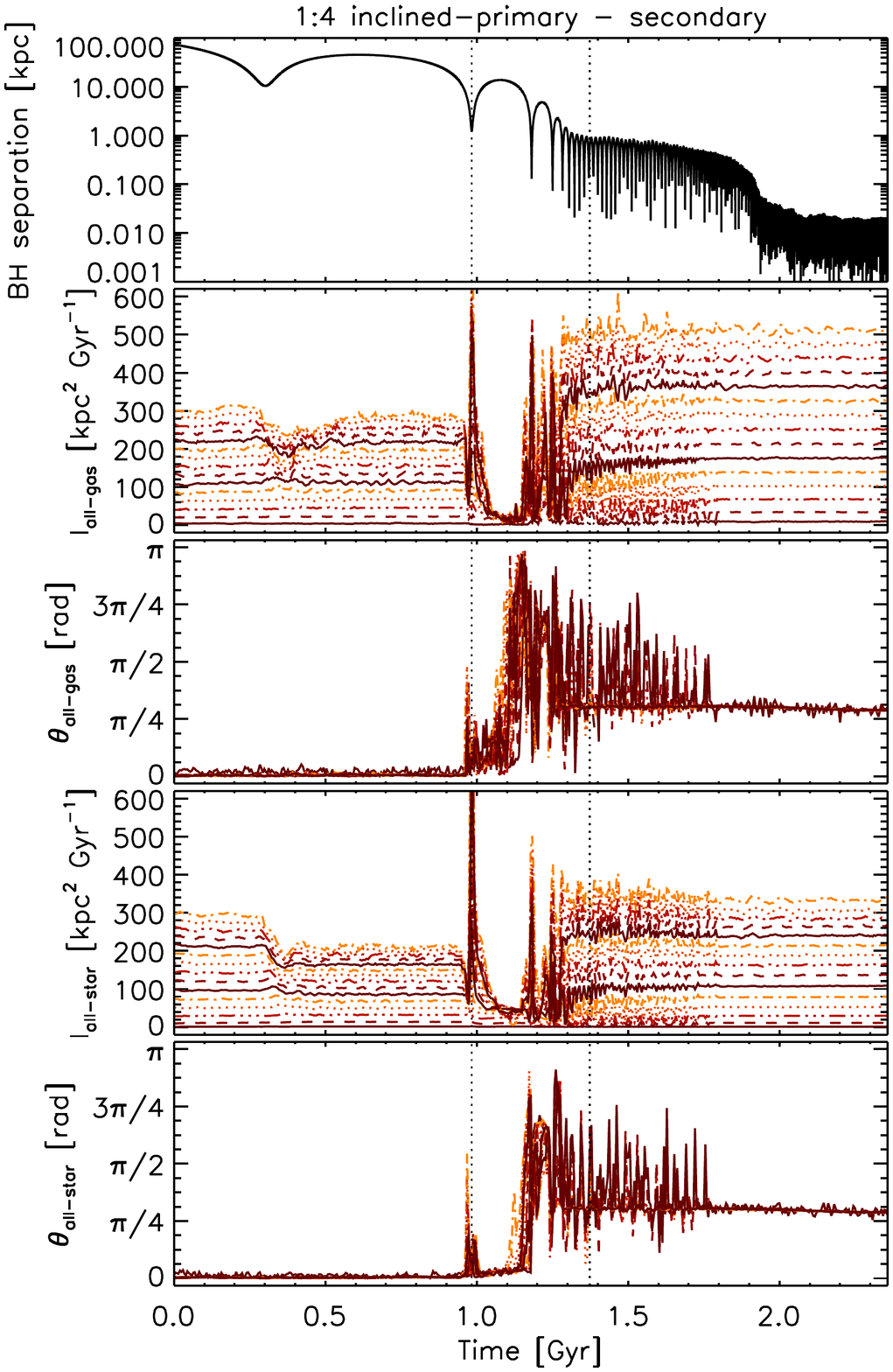}
\hspace{50.0pt}\includegraphics[trim = 11mm 14mm 30mm 7mm, clip, width=0.8\columnwidth,angle=0]{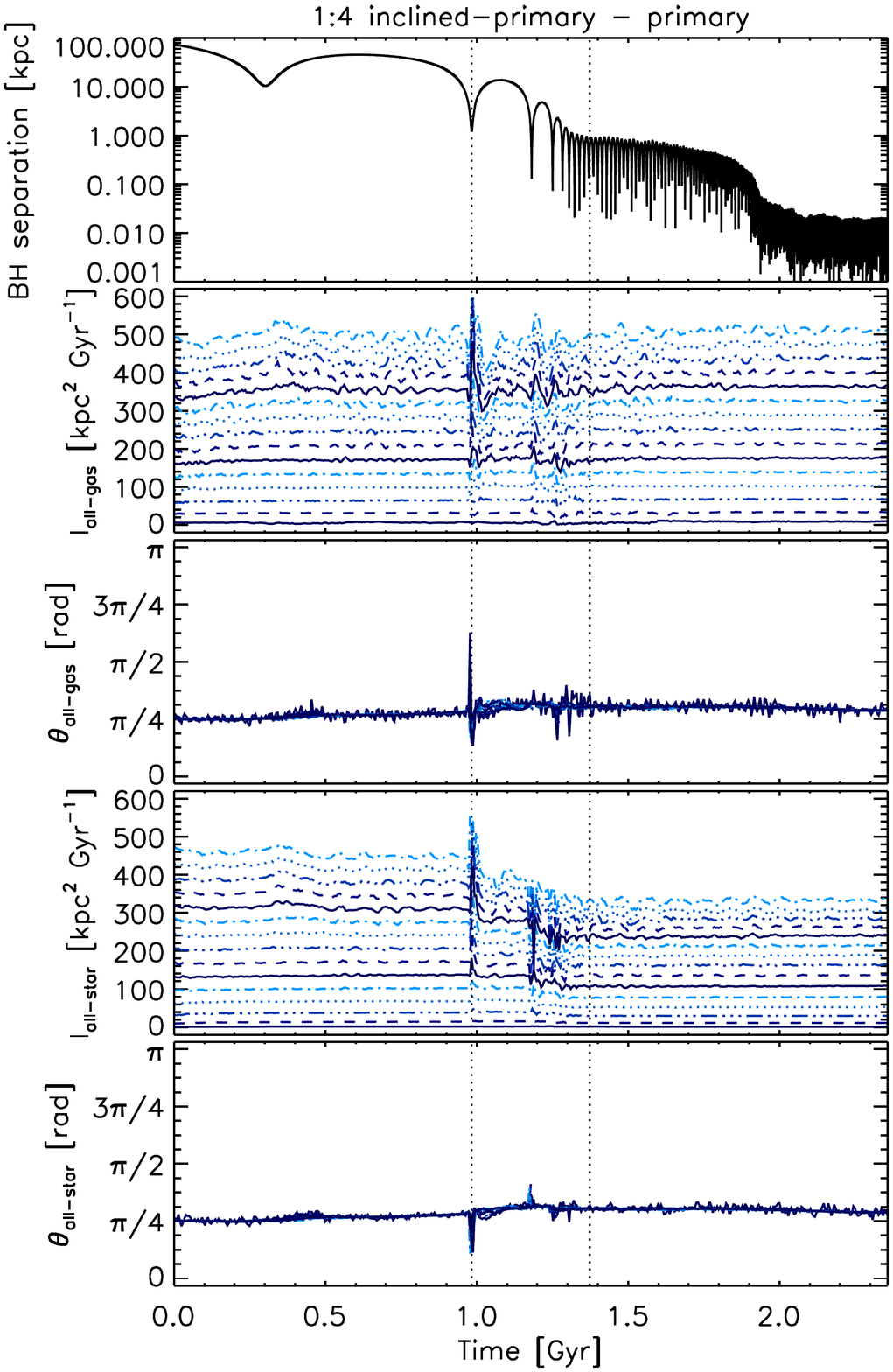}
\vspace{-5pt}
\caption[]{Temporal evolution of the specific angular momentum for the secondary (left) and primary (right) galaxy in the 1:4 inclined-primary merger. Same as Fig.~1 in the main text.}
\label{angmomflips:fig:m4_hr_gf0_3_BHeff0_001_phi045000_angular_momentum_allgas_allstars_3kpc}
\end{figure*}

\begin{figure*}
\centering
\vspace{2.5pt}
\includegraphics[trim = 11mm 14mm 30mm 7mm, clip, width=0.8\columnwidth,angle=0]{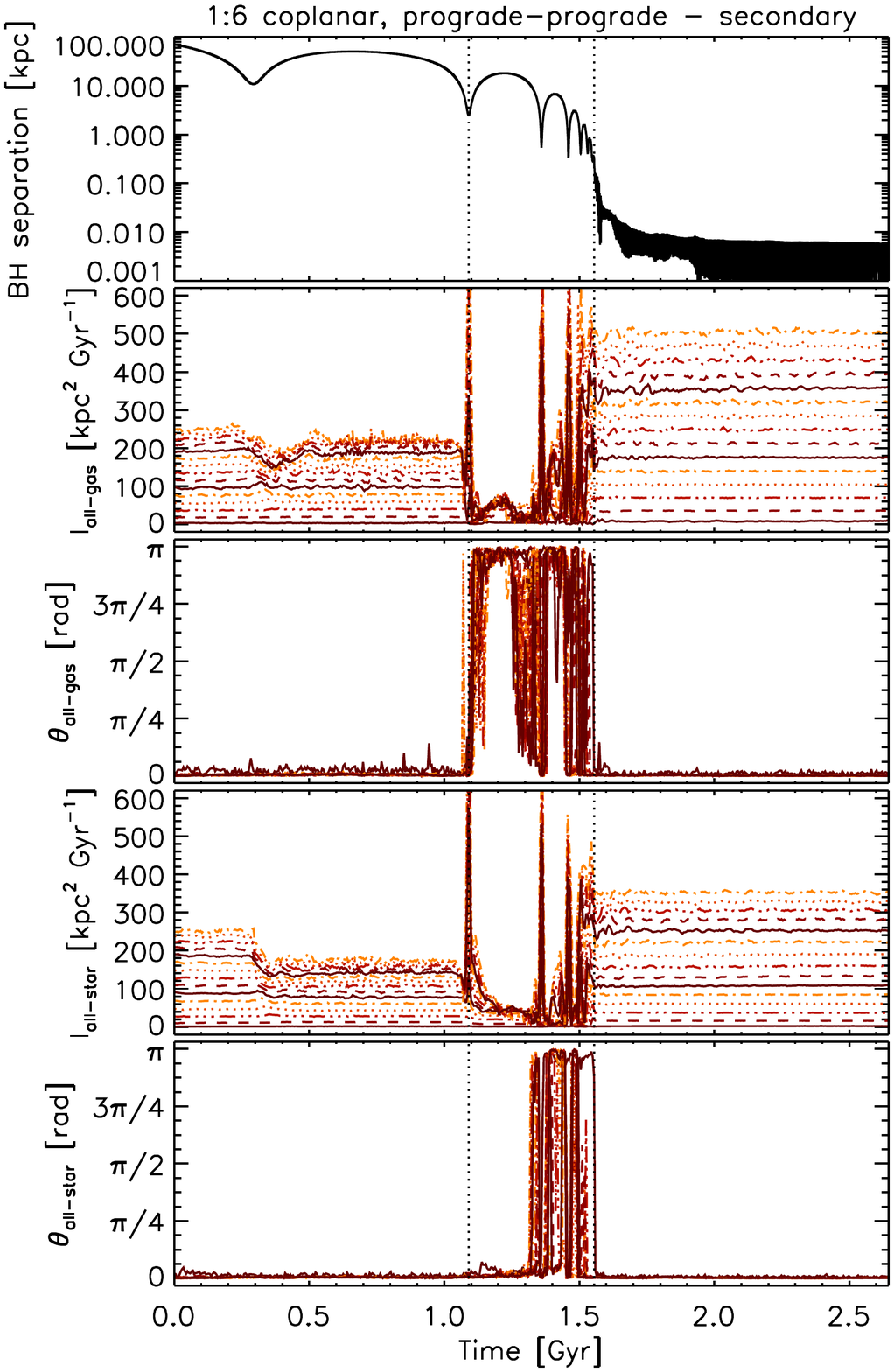}
\hspace{50.0pt}\includegraphics[trim = 11mm 14mm 30mm 7mm, clip, width=0.8\columnwidth,angle=0]{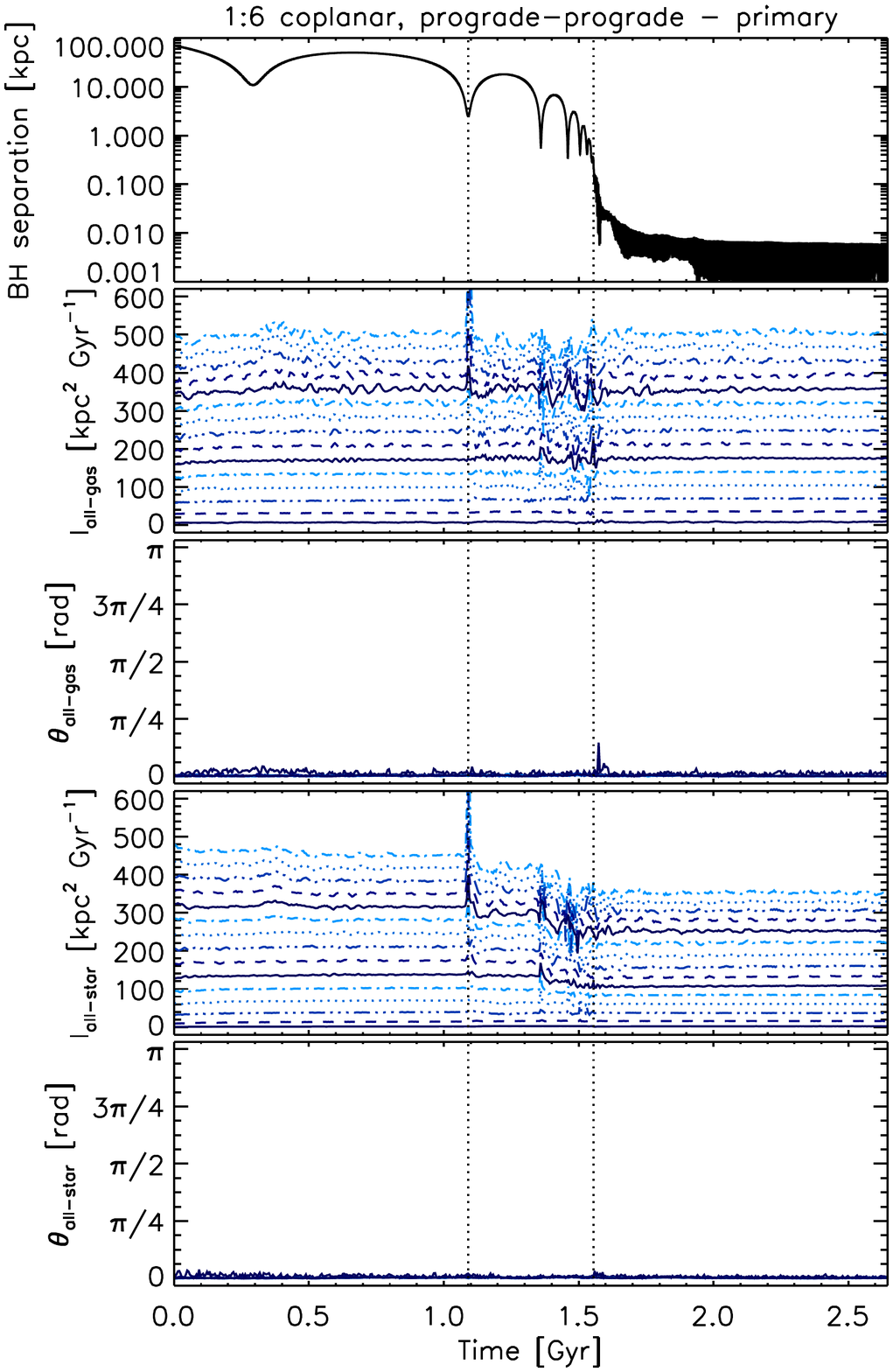}
\vspace{-5pt}
\caption[]{Temporal evolution of the specific angular momentum for the secondary (left) and primary (right) galaxy in the 1:6 coplanar, prograde--prograde merger. Same as Fig.~1 in the main text.}
\label{angmomflips:fig:m6_hr_gf0_3_BHeff0_001_phi000000_angular_momentum_allgas_allstars_3kpc}
\end{figure*}

\begin{figure*}
\centering
\vspace{2.5pt}
\includegraphics[trim = 11mm 14mm 30mm 7mm, clip, width=0.8\columnwidth,angle=0]{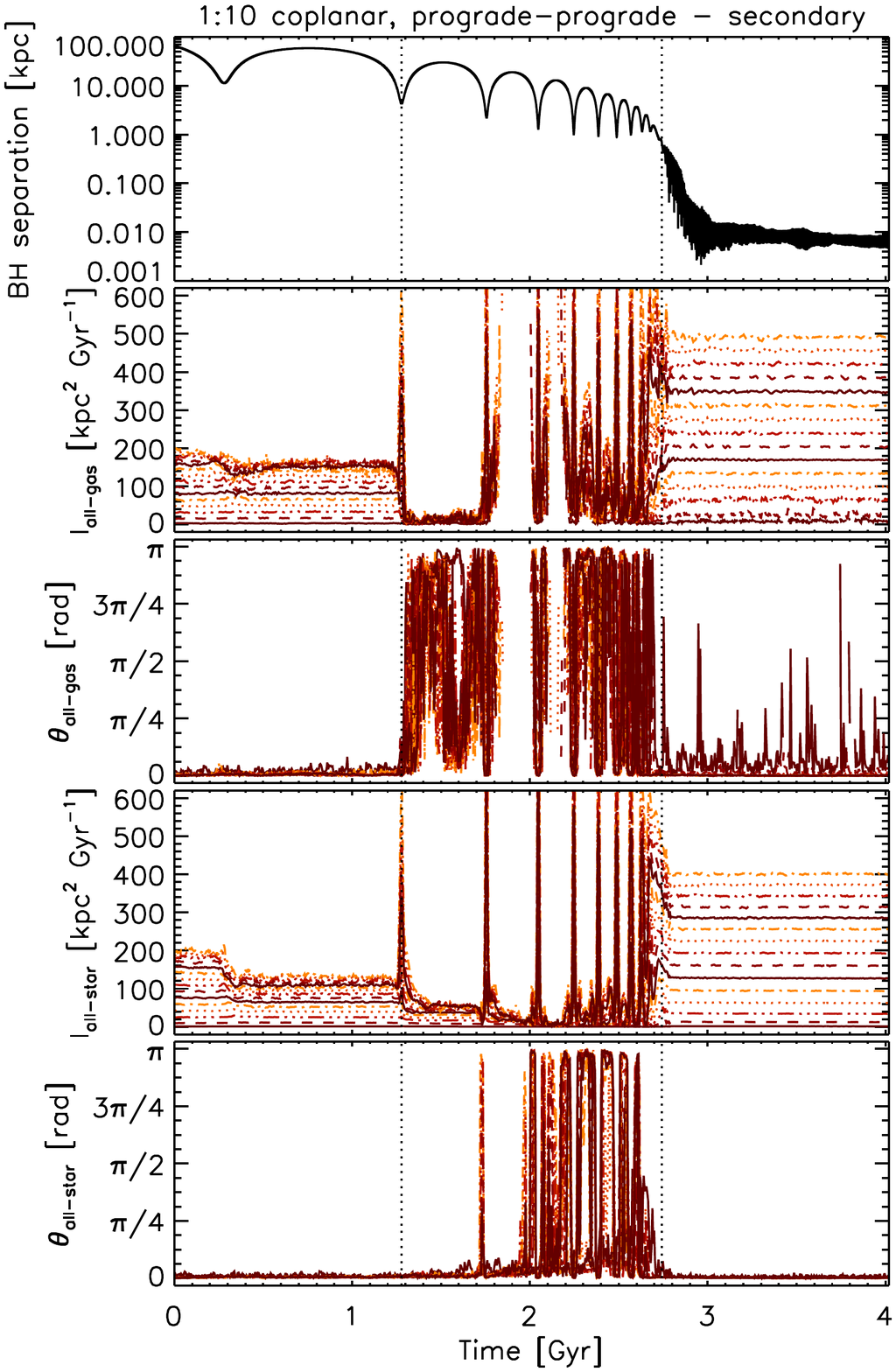}
\hspace{50.0pt}\includegraphics[trim = 11mm 14mm 30mm 7mm, clip, width=0.8\columnwidth,angle=0]{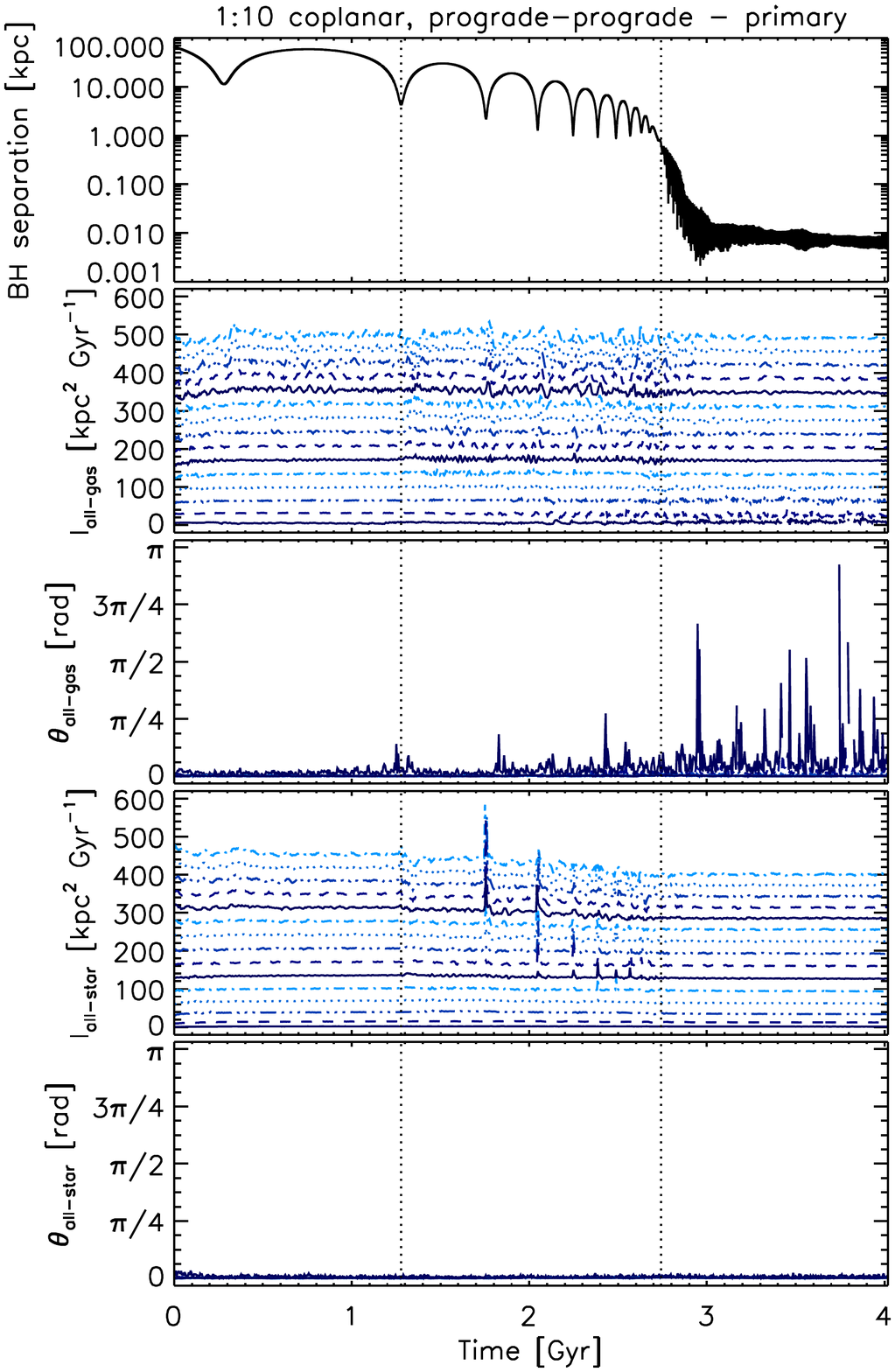}
\vspace{-5pt}
\caption[]{Temporal evolution of the specific angular momentum for the secondary (left) and primary (right) galaxy in the 1:10 coplanar, prograde--prograde merger. Same as Fig.~1 in the main text.}
\label{angmomflips:fig:m10_hr_gf0_3_BHeff0_001_phi000000_angular_momentum_allgas_allstars_3kpc}
\end{figure*}

\begin{figure*}
\centering
\vspace{2.5pt}
\includegraphics[trim = 11mm 14mm 30mm 7mm, clip, width=0.8\columnwidth,angle=0]{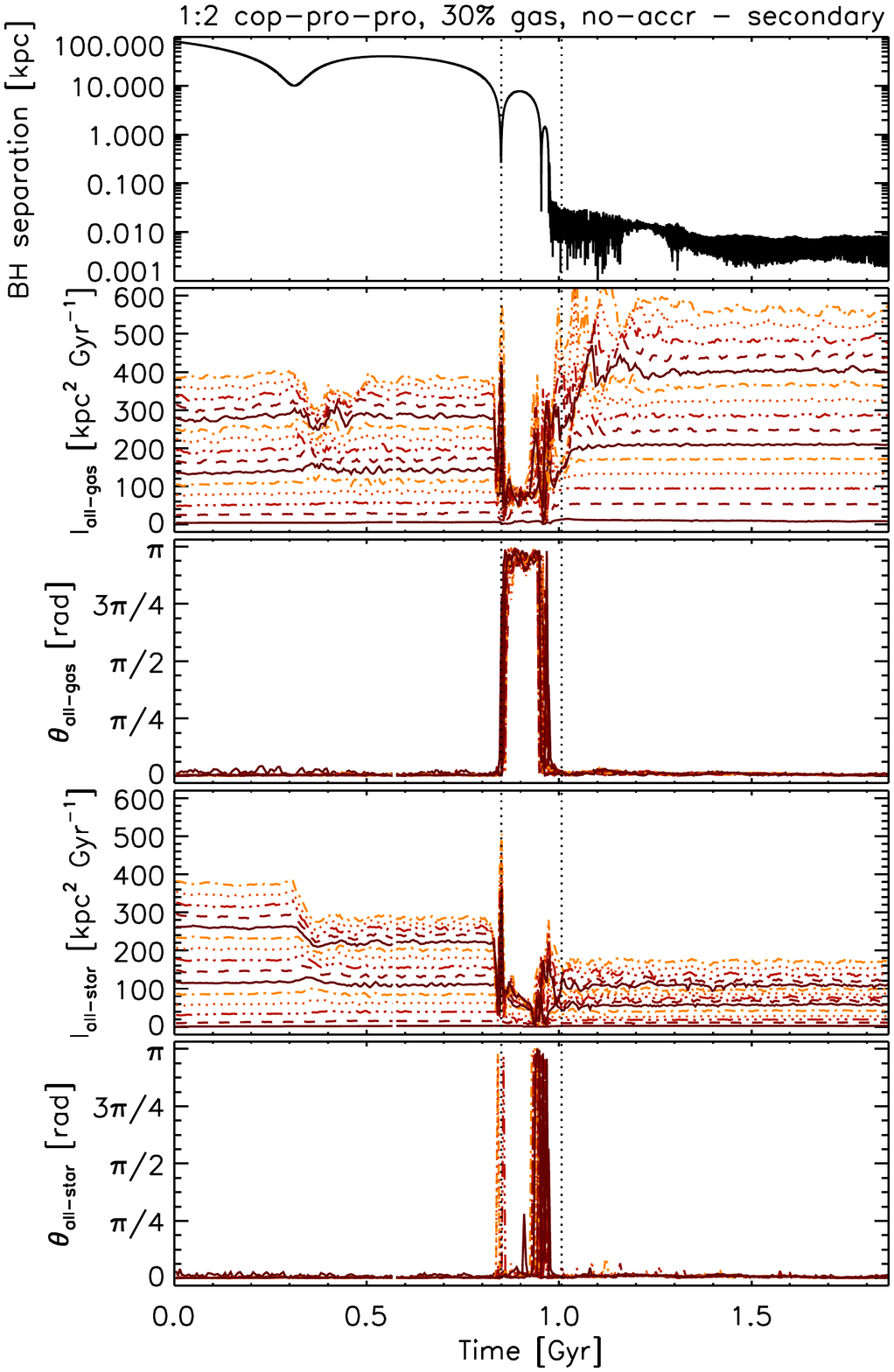}
\hspace{50.0pt}\includegraphics[trim = 11mm 14mm 30mm 7mm, clip, width=0.8\columnwidth,angle=0]{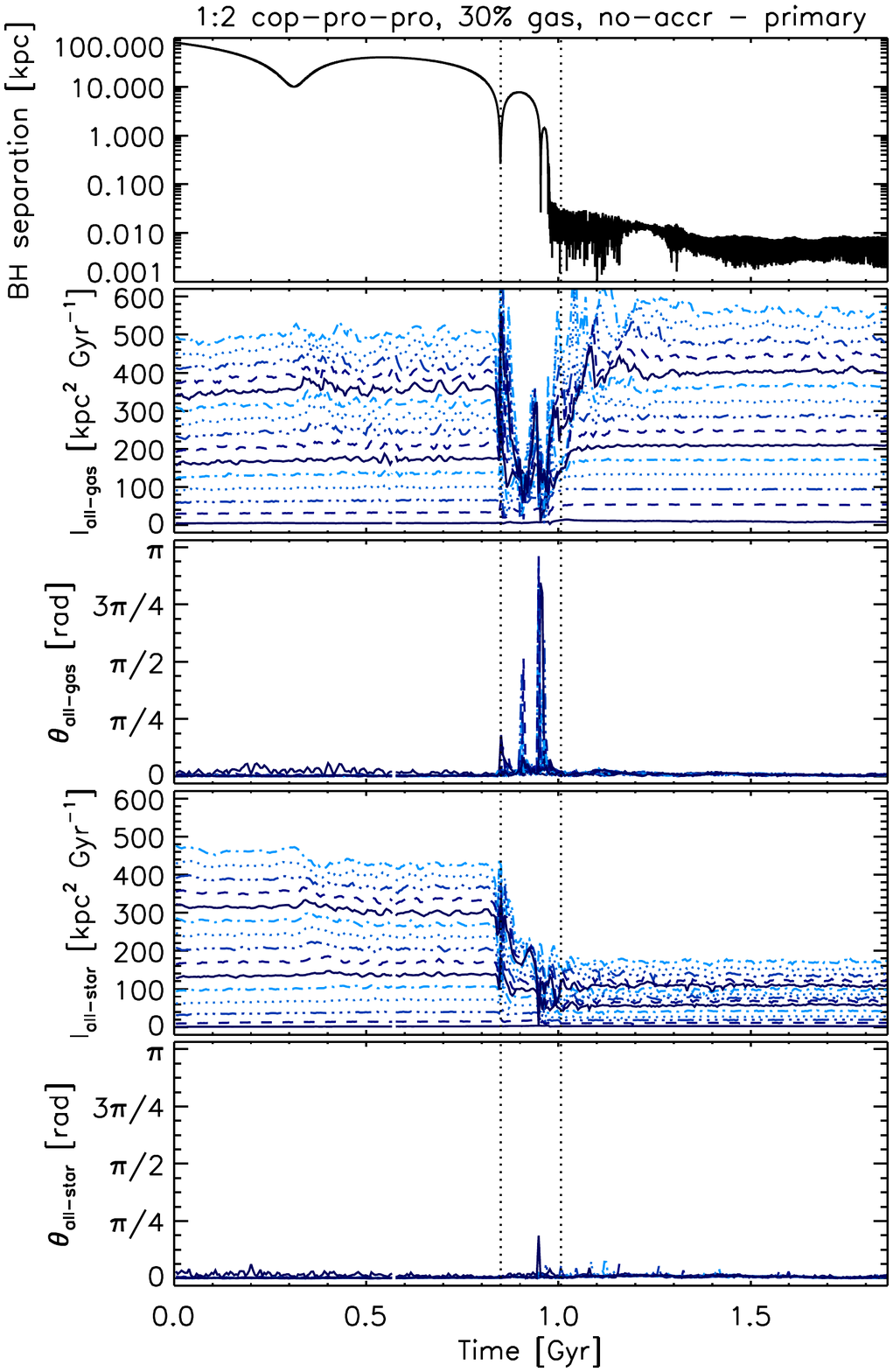}
\vspace{-5pt}
\caption[]{Temporal evolution of the specific angular momentum for the secondary (left) and primary (right) galaxy in the 1:2 coplanar, prograde--prograde merger with 30 per cent gas fraction and no BH accretion. Same as Fig.~1 in the main text.}
\label{angmomflips:fig:m2_hr_gf0_3_BHalpha0_0_phi000000_angular_momentum_allgas_allstars_3kpc}
\end{figure*}

\begin{figure*}
\centering
\vspace{2.5pt}
\includegraphics[trim = 11mm 14mm 30mm 7mm, clip, width=0.8\columnwidth,angle=0]{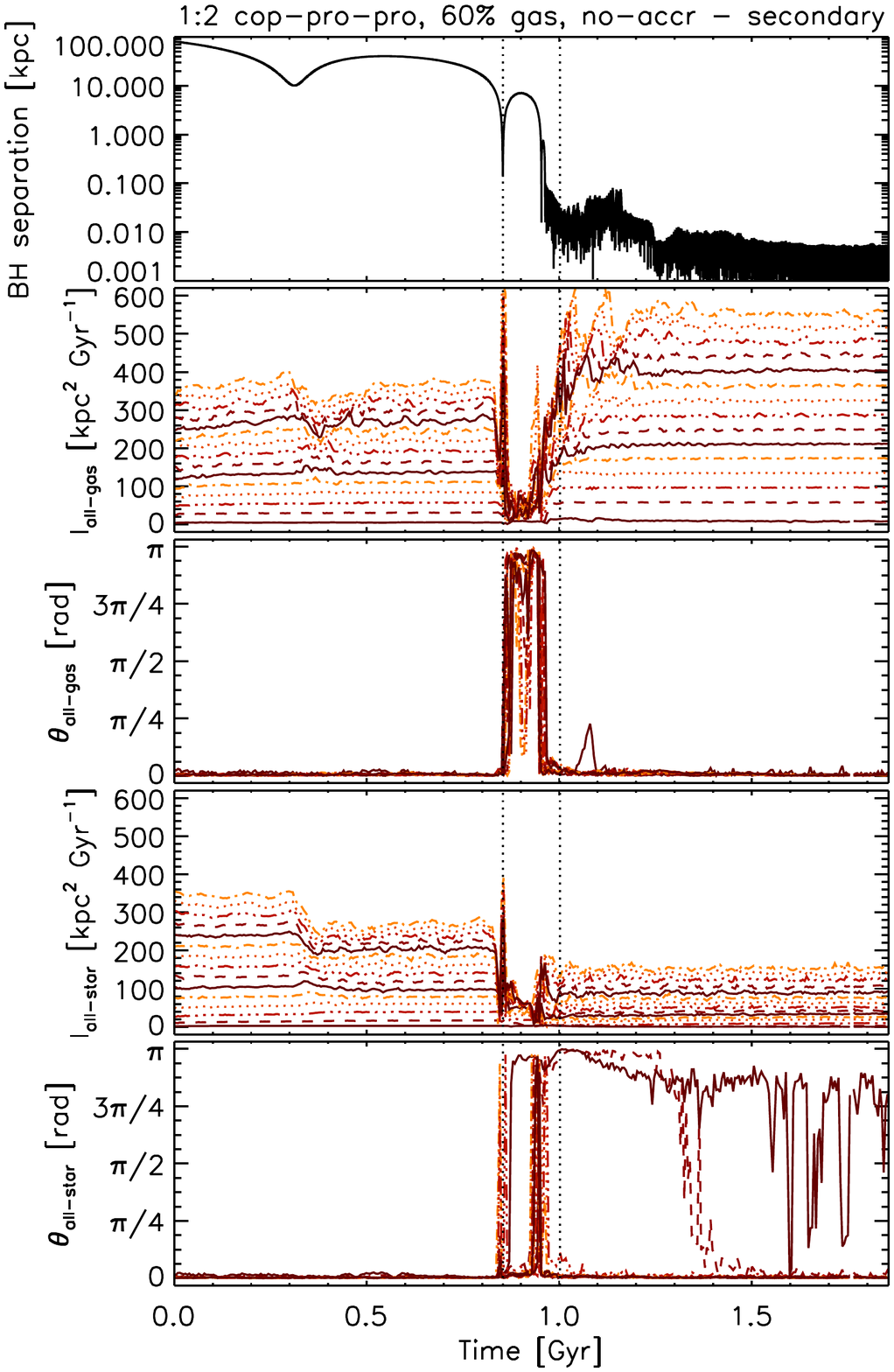}
\hspace{50.0pt}\includegraphics[trim = 11mm 14mm 30mm 7mm, clip, width=0.8\columnwidth,angle=0]{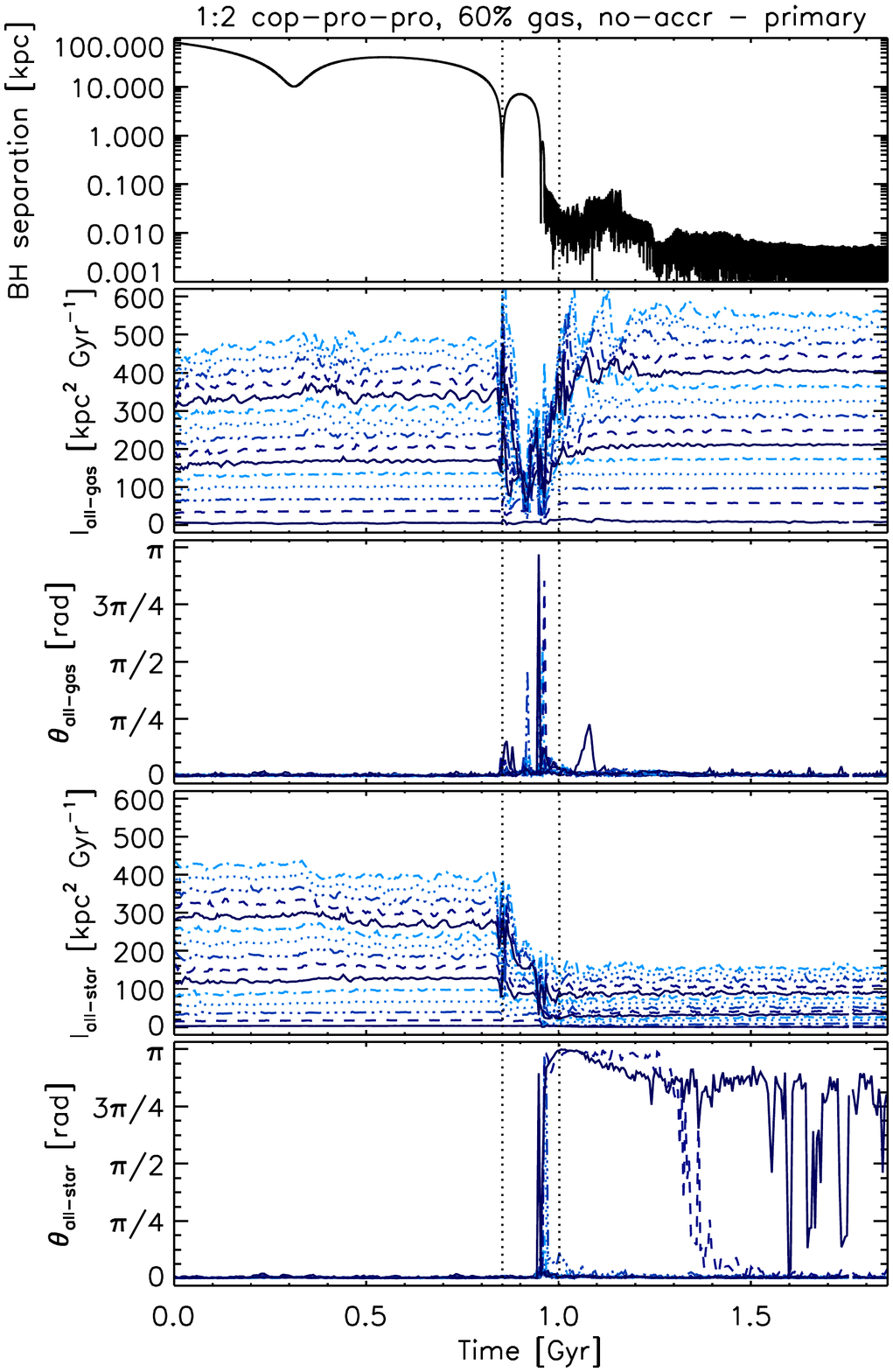}
\vspace{-5pt}
\caption[]{Temporal evolution of the specific angular momentum for the secondary (left) and primary (right) galaxy in the 1:2 coplanar, prograde--prograde merger with 60 per cent gas fraction and no BH accretion. Same as Fig.~1 in the main text.}
\label{angmomflips:fig:m2_hr_gf0_6_BHalpha0_0_phi000000_angular_momentum_allgas_allstars_3kpc}
\end{figure*}

\begin{figure*}
\centering
\vspace{2.5pt}
\includegraphics[trim = 11mm 14mm 30mm 7mm, clip, width=0.8\columnwidth,angle=0]{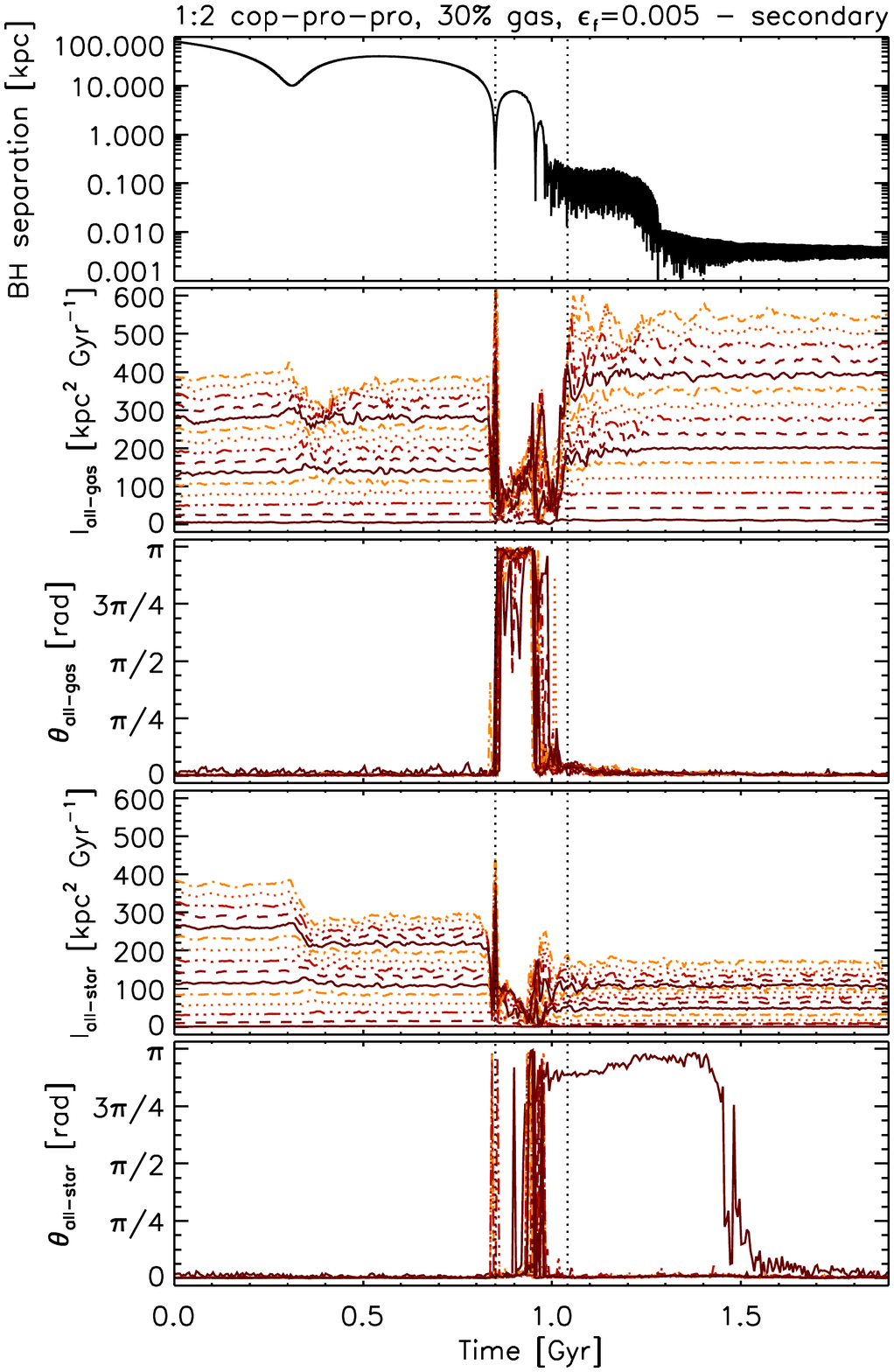}
\hspace{50.0pt}\includegraphics[trim = 11mm 14mm 30mm 7mm, clip, width=0.8\columnwidth,angle=0]{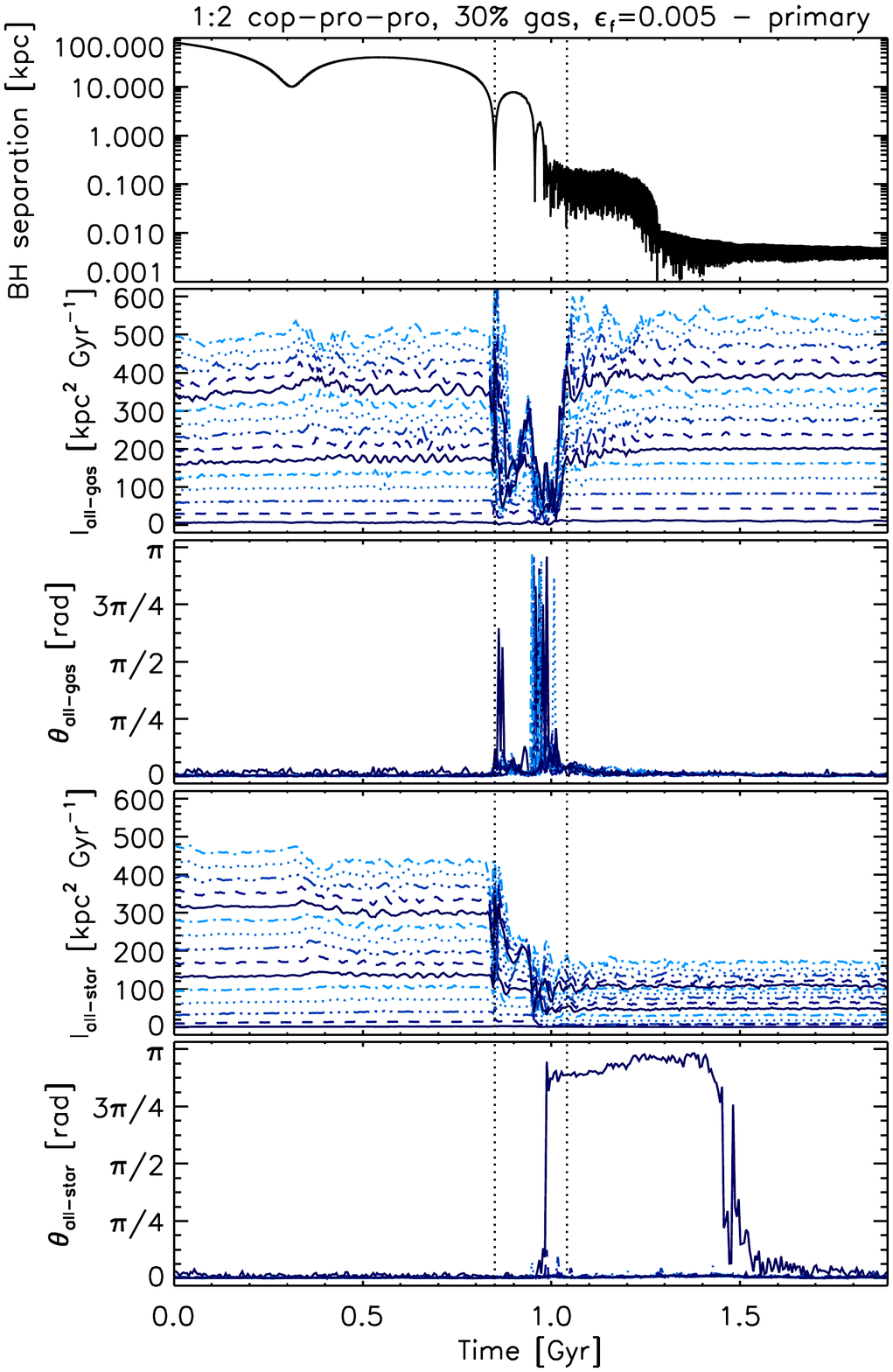}
\vspace{-5pt}
\caption[]{Temporal evolution of the specific angular momentum for the secondary (left) and primary (right) galaxy in the 1:2 coplanar, prograde--prograde merger with 30 per cent gas fraction and high BH feedback efficiency. Same as Fig.~1 in the main text.}
\label{angmomflips:fig:m2_hr_gf0_3_BHeff0_005_phi000000_angular_momentum_allgas_allstars_3kpc}
\end{figure*}

\begin{figure*}
\centering
\vspace{2.5pt}
\includegraphics[trim = 11mm 14mm 30mm 7mm, clip, width=0.8\columnwidth,angle=0]{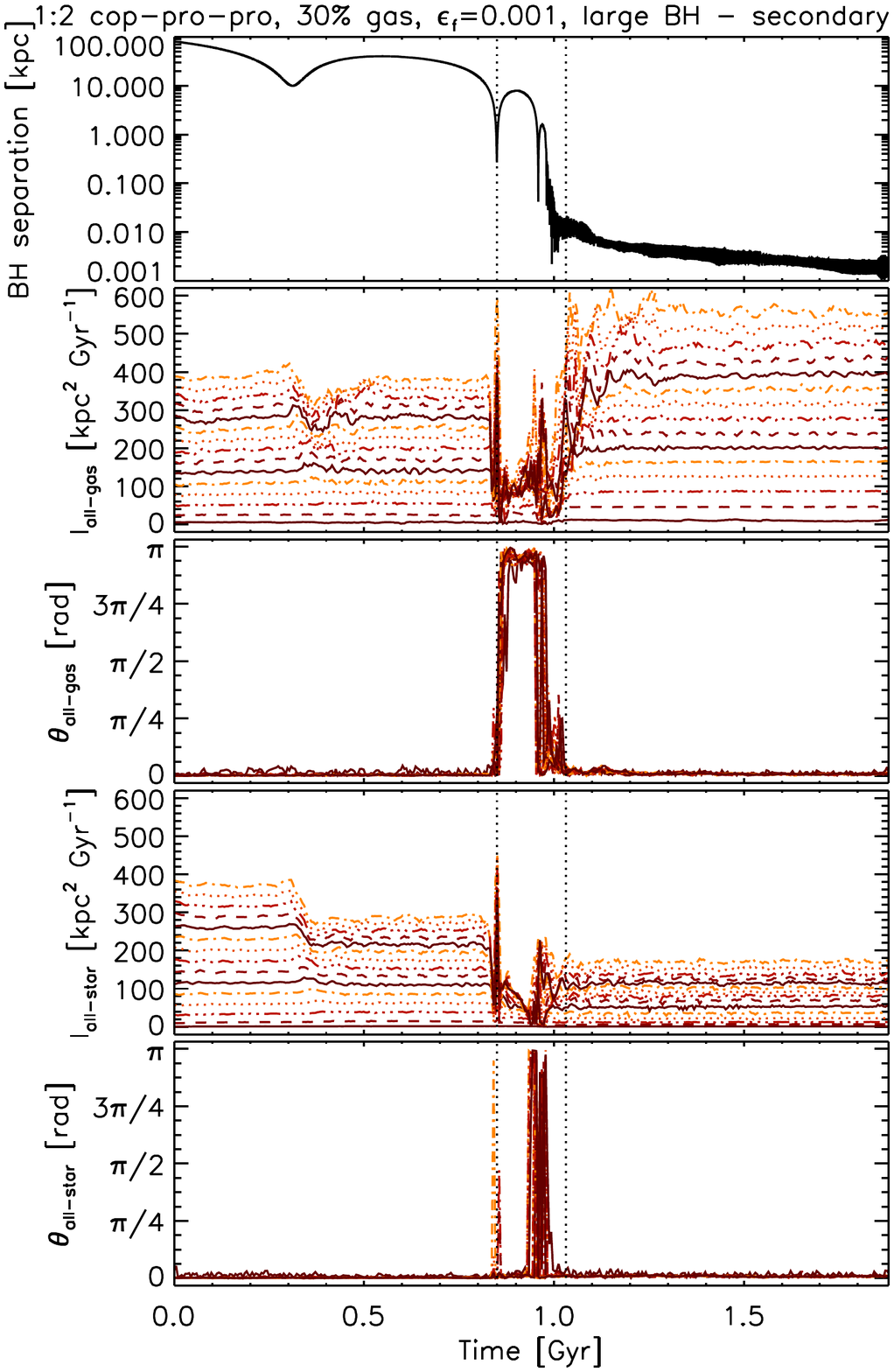}
\hspace{50.0pt}\includegraphics[trim = 11mm 14mm 30mm 7mm, clip, width=0.8\columnwidth,angle=0]{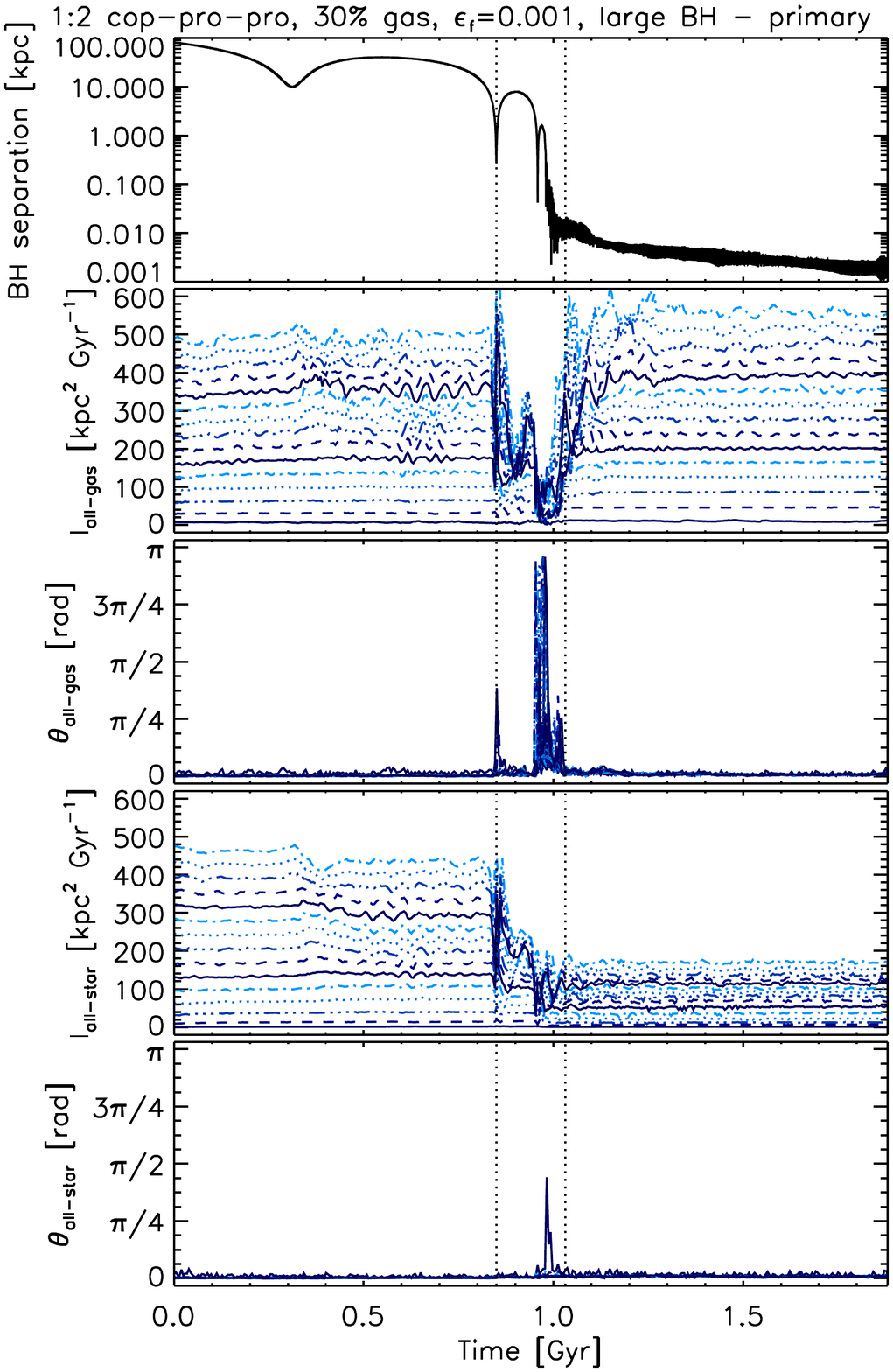}
\vspace{-5pt}
\caption[]{Temporal evolution of the specific angular momentum for the secondary (left) and primary (right) galaxy in the 1:2 coplanar, prograde--prograde merger with 30 per cent gas fraction, standard BH feedback efficiency, and larger-mass BHs. Same as Fig.~1 in the main text.}
\label{angmomflips:fig:m2_hr_gf0_3_BHeff0_001_phi000000_largerBH_angular_momentum_allgas_allstars_3kpc}
\end{figure*}

\end{document}